%% file: thin-film-electromechanics_article.tex
\documentclass[11pt,oneside,notitlepage]{article}

\usepackage[letterpaper, total={6.5in, 8.5in}]{geometry}

\usepackage{amsmath}
\usepackage{amssymb}
\usepackage{amsthm}
\usepackage{array}
\usepackage{bm}
\usepackage{mathrsfs}
\usepackage{mathtools}

\usepackage[font=small]{caption} 
\usepackage[font=small]{subcaption}

\usepackage{cite}
\usepackage{comment}
\usepackage{enumitem}
\usepackage{float}
\usepackage[T1]{fontenc}
\usepackage{graphicx}
\usepackage[
colorlinks=true,
citecolor=MidnightBlue,
linkcolor=BrickRed,
urlcolor=MidnightBlue,
breaklinks=true,
linktoc=all
]{hyperref}
\usepackage{listings}
\usepackage{multicol}
\usepackage{stmaryrd}
\usepackage{textgreek}
\usepackage{titlesec}
\usepackage{tocloft}
\usepackage{wrapfig}
\usepackage[usenames,dvipsnames]{xcolor}
\usepackage{xy}
\usepackage{amsmath}
\usepackage{relsize}
\usepackage{empheq}
\usepackage{comment}
\usepackage{booktabs}
\usepackage{layouts}
\usepackage{environ}
\usepackage{pdflscape}
\usepackage{afterpage}
\usepackage{placeins} 
\usepackage[normalem]{ulem} 

\usepackage{pgfplots}
\pgfplotsset{compat=1.15}
\usepackage{pgfmath}
\usetikzlibrary{patterns, calc, shapes.geometric, positioning, arrows.meta,backgrounds}
\usepackage{tikz-imagelabels}
\usepackage{tikz}

\newboolean{showcaption}
\setboolean{showcaption}{true}
\newcommand{\ifcap}[2]{\ifthenelse{\boolean{showcaption}}{#1}{#2}}

\def\imagebox#1#2{\vtop to #1{\vfill\null\hbox{#2}\vfill}}

\makeatletter
\newsavebox{\measure@tikzpicture}
\NewEnviron{scaletikzpicturetowidth}[1]{%
  \def\tikz@width{#1}%
  \begin{lrbox}{\measure@tikzpicture}%
  \BODY
  \end{lrbox}%
  \pgfmathparse{#1/\wd\measure@tikzpicture}%
  \BODY
}
\makeatother

\input{electromechanics-article/Content/custom-commands}

\titleformat{\paragraph}[hang]{\normalfont\normalsize\bfseries}{\theparagraph}{1em}{}
\titlespacing*{\paragraph}{0pt}{3.25ex plus 1ex minus .2ex}{0.5em}

\definecolor{dkgreen}{rgb}{0,0.6,0}
\definecolor{gray}{rgb}{0.5,0.5,0.5}
\definecolor{mauve}{rgb}{0.58,0,0.82}

\newcolumntype{A}{ >{$} r <{$} @{} >{${}} l <{$} } 

\allowdisplaybreaks

\renewcommand{\thefootnote}{\fnsymbol{footnote}}

\usepackage{sepfootnotes}

\makeatletter
\def\@fnsymbol#1{%
	\ensuremath{\ifcase#1\or
	\ddagger\or			
	\mathsection\or		
	*\or		
	\dagger\or			
	\mathflat\or				
	\|\or				
	\ddagger\ddagger\or	
	\dagger\dagger\or	
	**					
	\else\@ctrerr\fi}}
\makeatother

\pgfdeclarelayer{bg0}    
\pgfdeclarelayer{bg1}
\pgfsetlayers{bg0,bg1,main}  

\begin{document}

	\begin{center}
    
		{\textbf{
                {\large The $(2+\delta)$-dimensional theory of the electromechanics of lipid membranes: \\[4pt] II. Balance laws}
            }} \\

		\vspace{0.21in}

            \sepfootnotecontent{YO}{\href{mailto:yannick.omar@berkeley.edu}{yannick.omar\textit{@}berkeley.edu}}   
            \sepfootnotecontent{ZL}{\href{mailto:zlipel@berkeley.edu}{zlipel\textit{@}berkeley.edu}}
            \sepfootnotecontent{KM}{\href{mailto:kranthi@berkeley.edu}{kranthi\textit{@}berkeley.edu}}

		{\small
			Yannick A. D. Omar\textsuperscript{1,}\sepfootnote{YO}\,, Zachary G. Lipel\textsuperscript{1,}\sepfootnote{ZL}\,,
			and Kranthi K. Mandadapu\textsuperscript{1,2,}\sepfootnote{KM}\\
		}
		\vspace{0.25in}

		{\footnotesize
			{
				$^1$
				Department of Chemical \& Biomolecular Engineering,
				University of California, Berkeley, CA 94720, USA
				\\[3pt]
				$^2$
				Chemical Sciences Division, Lawrence Berkeley National Laboratory, CA 94720, USA
				\\
			}
		}
	\end{center}

	\vspace{13pt}
	%
	%

	\begin{abstract}
        This article is the second of a three-part series that derives a self-consistent theoretical framework of the electromechanics of arbitrarily curved lipid membranes. Existing continuum theories commonly treat lipid membranes as strictly two-dimensional surfaces. While this approach is successful in many purely mechanical applications, strict surface theories fail to capture the electric potential drop across lipid membranes, the effects of surface charges, and electric fields within the membrane. Consequently, they do not accurately resolve Maxwell stresses in the interior of the membrane and its proximity. Furthermore, surface theories are generally unable to capture the effects of distinct velocities and tractions at the interfaces between lipid membranes and their surrounding bulk fluids. To address these shortcomings, we apply a recently proposed dimension reduction method to the three-dimensional, electromechanical balance laws. This approach allows us to derive an effective surface theory without taking the limit of vanishing thickness, thus incorporating effects arising from the finite thickness of lipid membranes. We refer to this effective surface theory as $(2 + \delta)$-dimensional, where $\delta$ indicates the thickness. The resulting $(2 + \delta)$-dimensional equations of motion satisfy velocity and traction continuity conditions at the membrane-bulk interfaces, capture the effects of Maxwell stresses, and can directly incorporate three-dimensional constitutive models.
	\end{abstract}
	\vspace{15pt}

	%
	%

	{ \hypersetup{linkcolor=black} \tableofcontents }
	\vspace{20pt}

	%
	%

        \renewcommand{\thefootnote}{\arabic{footnote}}
        \setcounter{footnote}{0} 
	
        \input{electromechanics-article/Content/content.tex}

	%
	%

	\addcontentsline{toc}{section}{References}
	\bibliographystyle{electromechanics-article/bibliography/bibStyle}
	\bibliography{electromechanics-article/bibliography/bibliography}

\end{document}

%% file: electromechanics-article/Content/custom-commands.tex
\let\originalleft\left
\let\originalright\right
\newcommand{\leftR}{\mathopen{}\mathclose\bgroup\originalleft}
\newcommand{\rightR}{\aftergroup\egroup\originalright}

\newcommand{\diff}[1]{~\mathrm{d}#1}
\newcommand{\ddiff}[1]{\mathrm{d}#1}

\newcommand{\ChebT}[2]{P_{#1}\leftR(#2\rightR)}
\newcommand{\ChebTR}[1]{P_{#1}}

\newcommand{\divv}[1]{\mathrm{div}\leftR(#1\rightR)}

\newcommand{\bbar}[1]{\Bar{\Bar{#1}}}
\DeclareMathOperator{\tr}{tr}

\DeclareMathOperator*{\argmin}{arg\,min}

\newcommand{\KL}{K-L }

\newcommand{\SMsec}[1]{Sec.~{#1} of the SM}

\newcommand{\textspace}{\smallskip}

\tikzset{
    position/.style args={#1:#2 from #3}{
        at=(#3), xshift=#1, yshift=#2
    }
}

\newcommand{\bma}{\bm{a}}
\newcommand{\bmb}{\bm{b}}
\newcommand{\bmc}{\bm{c}}

\newcommand{\bmf}{\bm{f}}
\newcommand{\bmg}{\bm{g}}

\newcommand{\bmi}{\bm{i}}

\newcommand{\bml}{\bm{l}}
\newcommand{\bmm}{\bm{m}}
\newcommand{\bmn}{\bm{n}}

\newcommand{\bmp}{\bm{p}}

\newcommand{\bmt}{\bm{t}}

\newcommand{\bmv}{\bm{v}}

\newcommand{\bmx}{\bm{x}}

\newcommand{\bmA}{\bm{A}}

\newcommand{\bmC}{\bm{C}}
\newcommand{\bmD}{\bm{D}}
\newcommand{\bmE}{\bm{E}}
\newcommand{\bmF}{\bm{F}}
\newcommand{\bmG}{\bm{G}}

\newcommand{\bmN}{\bm{N}}

\newcommand{\bmT}{\bm{T}}

\newcommand{\bmX}{{\bm{X}}}

\newcommand{\Ga}{\alpha}
\newcommand{\Gb}{\beta}
\newcommand{\Gg}{\gamma}
\newcommand{\Gd}{\delta}

%% file: electromechanics-article/Content/content.tex
\section{Introduction}\label{sec:intro}

This series of articles presents a self-consistent theoretical framework for understanding the electromechanics of arbitrarily curved lipid membranes. In part I \cite{omar2023ES}, we introduced a new dimension reduction method to derive an \textit{effective} surface theory for the electrostatics of thin films. In this article, we focus on deriving dimensionally-reduced mechanical balance laws and equations of motions for thin films.\textspace 

Continuum theories describing the mechanical behavior of lipid membranes commonly treat them as strictly two-dimensional surfaces, an assumption motivated by their small thickness compared to their lateral dimensions. While this approach has proved successful in many applications \cite{helfrich1973elastic,servuss1976measurement,brochard1975frequency,seifert1997configurations,derenyi2002formation,sorre2009curvature, abreu2014fluid}, it is not suitable to describe the coupled electrical and mechanical behavior of lipid membranes. As described in part I \cite{omar2023ES}, strict surface theories fail to capture the potential drop across lipid membranes, the effects of surface charges, and the electric field within the membrane. Consequently, strict surface theories can also not resolve Maxwell stresses in the interior of the membrane and their proximity. Furthermore, the conditions of traction and velocity continuity at the two interfaces between lipid membranes and their surrounding bulk fluids ($\mathcal{S}^\pm$ in Fig.~\ref{fig:schematic_setup}) cannot be incorporated into surface theories. As a result, surface theories do not appropriately describe the effects of distinct velocities and tractions at the membrane-bulk interfaces.\textspace

Addressing the aforementioned shortcomings of strict surface theories in modeling the electromechanics of lipid membranes requires accounting for their finite thickness. The dimension reduction method introduced in Part I \cite{omar2023ES} allows deriving \textit{effective} surface theories without taking the limit of vanishing thickness, thus incorporating effects arising from the finite thickness of lipid membranes. 
We refer to these effective surface theories as $(2+\delta)$-dimensional, where $\delta$ indicates the thickness. The proposed method is based on low-order polynomial expansions of the solution along the thickness direction of the lipid membrane. Using this method, we can systematically derive an effective surface theory for the mechanical balance laws and equations of motion of lipid membranes. The resulting $(2+\delta)$-dimensional equations of motion satisfy the aforementioned conditions of velocity and traction continuity, capture the effects of Maxwell stresses, and can also directly incorporate three-dimensional constitutive models. Thus, this article contributes the second component required to establish a self-consistent framework for understanding the electromechanics of lipid membranes.\textspace

The remainder of this article is structured as follows. In Sec.~\ref{sec:3DMechanics_thin_films}, we describe the kinematics of thin films and reformulate the Kirchhoff-Love assumptions as constraints. We then briefly revisit the three-dimensional mechanical balance laws and boundary conditions for thin films. In Sec.~\ref{sec:dimred_method}, the dimension reduction method is summarized and all assumptions required for the derivation of the $(2+\delta)$-dimensional theory are introduced. Subsequently, we derive the $(2+\delta)$-dimensional balances of mass, angular, and linear momentum in Sec.~\ref{sec:dimred_balances}, which form the basis of the equations of motion presented in Sec.~\ref{sec:EOMs} along with the corresponding $(2+\delta)$-dimensional boundary conditions. 

\section{Three-Dimensional Mechanics of Thin Films} \label{sec:3DMechanics_thin_films}

This section introduces the three-dimensional continuum theory governing the electromechanics of thin films, formulated using a differential geometric framework, and serves as the starting point for deriving the dimensionally-reduced theory in Sec.~\ref{sec:dimred_balances}. We begin by describing the kinematics of thin films under the Kirchhoff-Love (K-L) assumptions and then reinterpret the \KL assumptions as kinematic constraints. Subsequently, we summarize the balance laws of three-dimensional, isothermal continuum mechanics.\textspace

\paragraph{Notation} Vectors and tensors are denoted by bold letters (e.g. $\bm{v}$) and matrices are denoted using square brackets (e.g. $\left[ B_{ij} \right]$). Subscripts and superscripts indicate covariant and contravariant components (e.g. $v_i, \, v^i$), respectively, and repeated sub- and superscripts imply Einstein's summation convention. Greek indices (e.g. $\alpha, \, \beta$) take values $\{1,2\}$ while Latin indices (e.g. $i,\,j$) take values $\{1,2,3\}$ except when used in the context of Chebyshev expansions. For some function $f$, the short-hand notations $f\leftR(\theta^\alpha\rightR)$ and $f\leftR(\theta^i\rightR)$ imply $f\leftR(\theta^1,\theta^2\rightR)$ and $f\leftR(\theta^1,\theta^2, \theta^3\rightR)$, respectively, where $\theta^1,\theta^2,\theta^3$ are curvilinear coordinates. Partial derivatives with respect to coordinates $\theta^k$ are denoted by a comma (e.g. $a_{i,k} = \partial a_i/\partial \theta^k$) and covariant derivatives of vector components $a^i$ and tensor components $A^{ij}$ in three-dimensional space are denoted by $a^i\bigr\rvert_k$ and $A^{ij}\bigr\rvert_{k}$, respectively. Similarly, the covariant derivatives of vector components $a^\alpha$ and tensor components $A^{\alpha\beta}$ defined on the tangent space of a surface in three dimensions are denoted by $a^\alpha_{:\gamma}$ and $A^{\alpha\beta}_{;\gamma}$, respectively. For non-symmetric, second-order tensors, it is necessary to indicate the order of indices, which is achieved here using a dot. For example, $A_{i}^{\,.\,j}$ implies that $i$ is the first index while $j$ is the second index. However, most tensors considered in this article are symmetric such that the order of co- and contravariant components is inconsequential and does not need to be indicated, i.e. $A_{i}^{\,.\,j} = A_{\,.\,i}^{j} =A_i^j\,$.

\subsection{Kinematics} \label{sec:kinematics}

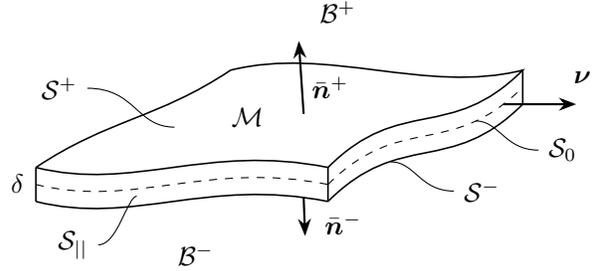
\begin{wrapfigure}{r}{0.49\textwidth}
    \centering
    \input{electromechanics-article/Figures/thin_shell_setup}
    \caption{Schematic of the setup used for our derivation. We consider a thin body $\mathcal{M}$ with thickness $\delta$ that separates the two bulk domains $\mathcal{B}^+$ and $\mathcal{B}^-$. }
    \label{fig:schematic_setup}
\end{wrapfigure}

We consider an oriented, thin film $\mathcal{M}$ with constant thickness $\delta$, embedded into a bulk domain (Fig.~\ref{fig:schematic_setup}). The segments of the bulk domain that lie above and below $\mathcal{M}$ are denoted by $\mathcal{B}^\pm$, respectively. The surfaces $\mathcal{S}^\pm$ form the interfaces between $\mathcal{M}$ and $\mathcal{B}^\pm$ with normal vectors $\bar{\bm{n}}^\pm$ pointing from $\mathcal{M}$ towards $\mathcal{B}^\pm$, and $\mathcal{S}_0$ denotes the mid-surface between $\mathcal{S}^+$ and $\mathcal{S}^-$. Finally, the lateral bounding surface is referred to as $\mathcal{S}_{||}$ and is equipped with the outward-pointing normal $\bm{\nu}$. \textspace

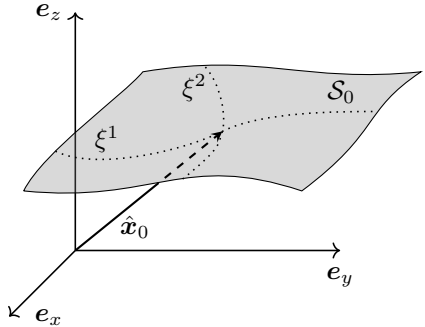
\begin{wrapfigure}{l}{0.4\textwidth}
    \centering
    \input{electromechanics-article/Figures/S0_parametrization}
    \caption{ The mid-surface $\mathcal{S}_0$ of the thin film $\mathcal{M}$ is parametrized using the curvilinear coordinates $\{\xi^1, \xi^2\} \in \Omega$.}
    \label{fig:S0_parametrization}
    \vspace{1cm}
\end{wrapfigure}

The kinematics and balance laws of the arbitrarily curved thin film are best described using a differential geometry framework \cite{aris2012vectors,itskov2019tensor}. To that end, we introduce a curvilinear parametrization $\left(\xi^1,\xi^2\right) \in \hat{\Omega}$ of the mid-surface $\mathcal{S}_0$ (Fig.~\ref{fig:S0_parametrization}) as well as a third parametric coordinate $\xi^3 \in \hat{\Xi}$ that parametrizes $\mathcal{M}$ through its thickness, with $\hat{\Omega}$ and $\hat{\Xi}$ denoting the respective parametric domains. The parametrization $\{\xi^i\}_{i=1,2,3}$ is independent of time $t$ and is therefore a Lagrangian or \textit{convected} parametrization \cite{rangamani2013interaction,sahu2020arbitrary}. In the following, we denote all quantities parametrized by $\{\xi^i\}_{i=1,2,3}$ by a \textit{hat} symbol.\textspace%

We introduce an additional curvilinear parametrization $\left(\theta^1,\theta^2\right) \in \Omega$ of the mid-surface $\mathcal{S}_0$ together with a third parametric coordinate $\theta^3 \in \Xi$ that parametrizes $\mathcal{M}$ through its thickness, where $\Omega$ and $\Xi$ are the respective parametric domains. As opposed to $\left\{\xi^i\right\}_{\alpha = 1,2,3}$, the parametrization $\{\theta^i\}_{i = 1,2,3}$ is considered to be Eulerian or \textit{surface-fixed} \cite{sahu2020arbitrary}, that is $\theta^i = \theta^i\leftR(\xi^j,t\rightR)$. The notion of an Eulerian parametrization will be made more explicit with the introduction of the mid-surface velocity below. As opposed to the Lagrangian parametrization, quantities parametrized by $\left\{\theta^i\right\}_{i=1,2,3}$ do not carry a distinct symbol. The introduction of both Lagrangian and Eulerian parametrizations is motivated by the viscous-elastic material behavior of lipid membranes: A Lagrangian parametrization makes it convenient to describe the elastic response to out-of-plane bending while an Eulerian parametrization is suitable to capture the in-plane fluidity of lipid membranes.\textspace

The position vector of a point on the thin film $\mathcal{M}$ is denoted by $\bm{x}\leftR(\theta^i\leftR(\xi^j,t\rightR),t\rightR) = \hat{\bm{x}}\leftR(\xi^i,t\rightR)$ while the position vector of a point on the mid-surface $\mathcal{S}_0$ is denoted by $\bm{x}_0\leftR(\theta^\alpha\leftR(\xi^\beta,t\rightR),t\rightR) = \hat{\bm{x}}_0\leftR(\xi^\alpha,t\rightR)$, as shown in Fig.~\ref{fig:S0_parametrization}. Furthermore, $\bm{x}\leftR(\theta^i,t\rightR)$ gives rise to $\bm{x}_0\leftR(\theta^i,t\rightR)$ through the condition $\bm{x}_0\leftR(\theta^1,\theta^2,t\rightR) = \bm{x}\leftR(\theta^1,\theta^2,\theta^3 = 0, t\rightR)$. We can now use the parametrizations to introduce various kinematics quantities. The mid-plane parametrization $\left\{\theta^\alpha\right\}_{\alpha=1,2}$ induces an in-plane, covariant basis $\{\bm{a}_\alpha\}_{\alpha=1,2}$ on $\mathcal{S}_0$, defined by 
\begin{align}
    \bm{a}_\alpha = \frac{\partial \bm{x}_0}{\partial \theta^\alpha}\biggr\lvert_{t}~. \label{eq:a_a_def}
\end{align}
This basis allows defining the normal vector to $\mathcal{S}_0$ as $\bm{n} = \bm{a}_1 \times \bm{a}_2/||\bm{a}_1 \times \bm{a}_2||$. 
Analogously to Eq.~\eqref{eq:a_a_def}, we also define the covariant basis $\{\bm{g}_i\}_{i=1,2,3}$ on $\mathcal{M}$ as
\begin{align}
    \bm{g}_i = \frac{\partial \bm{x}}{\partial \theta^i}\biggr\lvert_{t}~. \label{eq:g_i_def}
\end{align}
The covariant metric tensor components corresponding to Eqs.~\eqref{eq:a_a_def} and~\eqref{eq:g_i_def} are respectively defined as 
\begin{align}
    {a}_{\alpha\beta} &= \bm{a}_\alpha \cdot \bm{a}_\beta~, \label{eq:a_ab_def}\\
    g_{ij} &= \bm{g}_i \cdot \bm{g}_j~, \label{eq:g_ij_def}
\end{align}
and their inverses $\left[a_{\alpha\beta}\right]^{-1}=\left[a^{\alpha\beta}\right]$ and $\left[g_{ij}\right]^{-1}=\left[g^{ij}\right]$ provide the contravariant components of the metric tensors. The contravariant basis vectors can then be computed using the relations
\begin{align}
    \bm{a}^\alpha &= a^{\alpha\beta} \bm{a}_\beta~, \label{eq:aa_def}\\
    \bm{g}^i &= g^{ij} \bm{g}_j~. \label{eq:gi_def}
\end{align}
This allows us to express the metric tensor on the tangent plane to $\mathcal{S}_0$ as 
\begin{align}
    \bm{i} = \bm{a}_\alpha \otimes \bm{a}^\alpha~,
\end{align}
and the three-dimensional metric or identity tensor as
\begin{align}
    \bm{1} = \bm{g}_i \otimes \bm{g}^i~. \label{eq:I3_def}
\end{align}

The metric tensor components in Eqs.~\eqref{eq:a_ab_def} and~\eqref{eq:g_ij_def} and their inverses enable us to transform between covariant and contravariant components analogous to Eqs.~\eqref{eq:aa_def} and~\eqref{eq:gi_def}. To illustrate this, consider an arbitrary vector of the form $\bmc = c_\Ga \, \bma^\Ga = c^\Ga \, \bma_\Ga$ and an arbitrary tensor of the form $\bmA = A_{\Ga\Gb} \, \bma^\Ga \otimes \bma^\Gb = A_{\Ga}^{\,.\,\Gb} \, \bma^\Ga \otimes \bma_\Gb =  A_{\,.\,\Gb}^{\Ga} \, \bma_\Ga \otimes \bma^\Gb = A^{\Ga\Gb} \, \bma_\Ga \otimes \bma_\Gb$. The covariant, contravariant, and mixed components are then related by
\begin{align}
    c^\Ga &= a^{\Ga\Gb} \, c_\Gb~, \\
    c_\Ga &= a_{\Ga\Gb} \, c^\Gb~, \\
    A^{\Ga\Gb} &= a^{\Ga\Gg} A_{\Gg}^{\,.\,\Gb} = a^{\Gg\Gb} A_{\,.\,\Gg}^{\Ga} =  a^{\Ga\Gg} \, a^{\lambda\Gb} A_{\Gg\lambda}~, \\
    A_{\Ga\Gb} &= a_{\Ga\Gg} \, A_{\,.\,\Gb}^{\Gg} = a_{\Gg\Gb} \, A_{\Ga}^{\,.\,\Gg} = a_{\Ga\Gg} \, a_{\lambda\Gb} A^{\Gg\lambda}~.
\end{align}
Similarly, we can consider arbitrary vectors and tensors of the form
$\bmc = c_i \, \bmg^i = c^i \, \bmg_i$ and $\bmA = A_{ij} \, \bmg^i \otimes \bmg^j = A_{i}^{\,.\,j} \, \bmg^i \otimes \bmg_j =  A_{\,.\,j}^{i} \, \bmg_i \otimes \bmg^j = A^{ij} \, \bmg_i \otimes \bmg_j$. In this case, the various components are related by 
\begin{align}
    c^i &= g^{ij} \, c_j~, \\
    c_i &= g_{ij} \, c^j~, \\
    A^{ij} &= g^{ik} A_{k}^{\,.\,j} = g^{kj} A_{\,.\,k}^{i} =  g^{ik} \, g^{lj} A_{kl}~, \\
    A_{ij} &= g_{ik} \, A_{\,.\,j}^{k} = g_{kj} \, A_{i}^{\,.\,k} = g_{ik} \, g_{lj} A^{kl}~.
\end{align}

Before proceeding, we introduce an additional basis that is convenient to use when applying the dimension reduction method proposed in part 1 \cite{omar2023ES}. To this end, first recall that the basis vectors $\left\{\bmg_i\right\}_{i=1,2,3}$ form a basis of $\mathbb{R}^3$ that varies along $\theta^3$. In contrast, the basis vectors $\left\{\bma_\Ga\right\}_{\Ga=1,2}$ are independent of $\theta^3$ but do not form a basis of $\mathbb{R}^3$. However, we can introduce a new basis set $\left\{\check{\bmg}_i\right\}_{i=1,2,3}$, defined as 
\begin{align}
    \left\{\check{\bmg}_i\right\}_{i=1,2,3} = \{\bm{a}_1, \bm{a}_2, \bm{n}\}~, \label{eq:checkg_def}
\end{align}
that forms a basis of $\mathbb{R}^3$ and is independent of $\theta^3$. However, note that there does generally not exist a parametrization that induces the basis $\left\{\check{\bmg}_i\right\}_{i=1,2,3}$ through Eq.~\eqref{eq:g_i_def}, as will become apparent with the introduction of \KL kinematics below.\textspace 

Next, we define the curvature tensor of the mid-surface,
\begin{align}
    \bm{b} = b_{\alpha\beta} \, \bm{a}^\alpha \otimes \bm{a}^\beta = \left(\bm{a}_{\alpha,\beta} \cdot \bm{n} \right) \, \bm{a}^\alpha \otimes \bm{a}^\beta~. \label{eq:b_ab_def}
\end{align}
as well as the mean curvature $H$ and Gaussian curvature $K$ of the mid-surface,
\begin{align}
    H &= \frac{1}{2}\tr{\bm{b}}~, \label{eq:H_def} \\
    K &= \det{\bm{b}}~, \label{eq:K_def}
\end{align} 
where $\tr$ and $\det$ are the trace and determinant, respectively. The curvature tensor can also be written in terms of its spectral decomposition as
\begin{align}
    \bmb = \sum_{\Ga} \kappa_\Ga \bml_\Ga \otimes \bml^\Ga~, \label{eq:bmb_spectral_decomp}
\end{align}
with $\kappa_\Ga$ and $\bml_\Ga = \bml^\Ga$ denoting the principal curvatures and orthonormal eigenvectors, respectively. Equation~\eqref{eq:bmb_spectral_decomp} permits expressing the mean and Gaussian curvatures in terms of the principal curvatures as 
\begin{align}
    H &= \frac{1}{2}\left( \kappa_1 + \kappa_2\right)~, \label{eq:H_kappa}\\
    K &= \kappa_1 \kappa_2~. \label{eq:K_kappa}
\end{align}%
The results in Eqs.~\eqref{eq:bmb_spectral_decomp}--\eqref{eq:K_kappa} will prove useful in obtaining order of magnitude estimates in later sections.\textspace

We now note that Eqs.~\eqref{eq:a_a_def}--\eqref{eq:K_kappa} were introduced in terms of the Eulerian parametrization. However, they could be analogously introduced using the convected parametrization $\{\xi^i\}_{i=1,2,3}$, with all quantities carrying a \textit{hat} to denote the different parametrization, e.g. $\hat{\bm{g}}_{{i}}$. Throughout this article, the Lagrangian counterparts of  Eqs.~\eqref{eq:a_a_def}--\eqref{eq:K_kappa} are used without defining them explicitly.\textspace

We now introduce the assumption of Kirchhoff-Love (K-L) kinematics, a widely used model for describing elastic thin bodies. It assumes that any point along the normal to the mid-surface remains on the normal and maintains a constant distance from the mid-surface during deformation\footnote{\KL kinematics can be rigorously justified in some cases through the so-called scaling method (see Refs \cite{ciarlet1979justification,podio1989exact,podio1990constraint,ciarlet1980justification,miara1994justificationI,miara1994justificationII}). } \cite{ciarlet2000theory}. In the Lagrangian parametrization, this amounts to writing the position vector of a point on $\mathcal{M}$ as
\begin{align}
    \hat{\bm{x}}\leftR(\xi^i,t\rightR) = \hat{\bm{x}}_0\leftR(\xi^\alpha,t\rightR) + \hat{\bm{n}}\leftR(\xi^\alpha,t\rightR)\,\xi^3~, \label{eq:KL_Lagrange}
\end{align}
where we chose, without loss of generality, $\xi^3 \in (-\delta/2,\delta/2)$. From Eq.~\eqref{eq:KL_Lagrange}, it is apparent that $\xi^3$ parametrizes $\mathcal{M}$ perpendicular to the mid-surface and therefore, along the thickness.\textspace

Equation~\eqref{eq:KL_Lagrange} also implies that the velocity $\hat{\bm{v}}$ is at most linear along $\xi^3$, which follows immediately from
\begin{align}
    \hat{\bm{v}} = \hat{\bm{x}}_{,t} &= \frac{\partial \hat{\bm{x}}}{\partial t}\biggr\rvert_{\xi^i}  =\frac{\partial \hat{\bm{x}}_0}{\partial t}\biggr\rvert_{\xi^\Ga} + \frac{\partial \hat{\bmn}_0}{\partial t}\biggr\rvert_{\xi^\Ga} \xi^3~. \label{eq:vLagrangian_def}
\end{align}
Equation~\eqref{eq:vLagrangian_def} motivates the definition of the mid-surface velocity as
\begin{align}
    \hat{\bm{v}}_0\leftR(\xi^\alpha, t\rightR)  = \hat{\bm{x}}_{0,t}\leftR(\xi^\alpha, t\rightR)~, 
\end{align}
which can be split into the in-plane velocity $\hat{v}_0^{{\Ga}}$ and normal velocity $\hat{v}_0^{3}$ of the mid-surface as
\begin{align}
    \hat{\bm{v}}_0 = \hat{v}_0^{\hat{\Ga}}\bma_{{\Ga}} + \hat{v}_0^{3} \hat{\bmn}~. \label{eq:}
\end{align}
Similarly, we define the first-order velocity as $\hat{\bmv}_1 = \hat{\bmn}_{,t}$. By using the relations $\hat{\bm{n}}_{,t} \cdot \hat{\bm{n}} = 0$ and $\left(\hat{\bma}_{\Ga}\right)_{,t} = \hat{\bm{v}}_{0,\alpha}$, and Eq.~\eqref{eq:b_ab_def}, we then find
\begin{align}
    \hat{\bm{n}}_{,t} &= - \left(\hat{\bm{v}}_{0,\alpha} \cdot \hat{\bm{n}}\right) \hat{\bm{a}}^{\alpha} \\
    &= -\left( \hat{v}^{3}_{0,\alpha} + \hat{v}_0^{\lambda} \hat{b}_{\lambda\alpha} \right) \hat{\bm{a}}^\alpha~. \label{eq:nt_2}
\end{align}
Combining Eqs.~\eqref{eq:KL_Lagrange}--\eqref{eq:nt_2} then yields 
\begin{align}
    \hat{\bm{v}}\leftR(\xi^i,t\rightR) = \hat{\bm{x}}_{,t}\leftR(\xi^i,t\rightR) &= \hat{\bm{v}}_0\leftR(\xi^\alpha,t\rightR) + \hat{\bm{v}}_1\leftR(\xi^\alpha,t\rightR)\, \xi^3~\\
    &= \hat{\bm{v}}_0 - \left( \hat{v}^{3}_{0,\alpha} + \hat{v}_0^{\lambda} \hat{b}_{\lambda\alpha} \right)\hat{\bm{a}}^{\alpha} \, \xi^3~, \label{eq:v_hat}
\end{align}
where we omitted the explicit dependence on the parametrization and time in the last step. Equation~\eqref{eq:v_hat} shows that the velocity of any point in $\mathcal{M}$ is linearly dependent on $\xi^3$ and entirely determined by the velocity and curvature of the mid-surface.\textspace

Above, we utilized the \KL assumptions to determine the position and velocity vectors in the Lagrangian parametrization. However, the fluid-like behavior of lipid membranes is more conveniently described using the Eulerian parametrization $\left\{\theta^i\right\}_{i=1,2,3}$. In the Eulerian parametrization, the velocity vector is found by first choosing a suitable form of the position vector and subsequently ensuring that the velocity is independent of the choice of parametrization, i.e. $\hat{\bmv}\leftR(\xi^i,t\rightR) = \bmv\leftR(\theta^j\leftR(\xi^i,t\rightR),t\rightR)$. This procedure is discussed in more detail in \SMsec{1.1} and only the results are summarized here.\textspace

Choosing the position vector in Eulerian parametrization as
\begin{align}
    \bmx\leftR(\theta^i,t\rightR) = {\bmx}_0\leftR(\theta^\alpha,t\rightR) + \bmn\leftR(\theta^\alpha,t\rightR) \,  \theta^3~,  \label{eq:KL_Eulerian}
\end{align}
is found to be suitable as it describes the same body $\mathcal{M} = \left\{ \hat{\bm{x}}\leftR(\xi^i,t\rightR):\left(\xi^1,\xi^2,\xi^3\right) \in \hat{\Omega} \times \hat{\Xi}\right\}$ as Eq.~\eqref{eq:KL_Lagrange} if $\theta^3 \in (-\delta/2,\delta/2)$. In \SMsec{1.1}, we show that Eq.~\eqref{eq:KL_Eulerian} together with $\hat{\bmv}\leftR(\xi^i,t\rightR) = \bmv\leftR(\theta^j\leftR(\xi^i,t\rightR),t\rightR)$ yields the following conditions on the Eulerian parametrization,
\begin{align}
    \theta^\Ga &= \theta^\Ga\leftR(\xi^\Gb,t\rightR)~,\label{eq:thetaa_xib}\\
    \dot{\theta}^3 &= 0~. \label{dot_theta3}
\end{align}
Equation~\eqref{eq:thetaa_xib} states that the Eulerian in-plane parametrization only depends on $\xi^\Gb$ and not on $\xi^3$. This ensures that the in-plane velocity varies at most linearly through the thickness. Likewise, Eq.~\eqref{dot_theta3} states that $\theta^3$ is independent of time, in turn implying that the normal velocity is constant through the thickness.\textspace

Equation~\eqref{dot_theta3} can be satisfied a-priori by making the \textit{choice} 
\begin{align}
    \theta^3\leftR(\xi^3,t\rightR) = \xi^3~. \label{eq:theta3_xi3}
\end{align}
Given Eq.~\eqref{eq:theta3_xi3}, the velocity vector in the Eulerian parametrization can be obtained by taking the material time derivative of Eq.~\eqref{eq:KL_Eulerian},
\begin{align}
    \bmv\leftR(\theta^i,t\rightR) &= \frac{\ddiff }{\ddiff t}\bmx_0\leftR(\theta^\Ga,t\rightR) + \frac{\ddiff }{\ddiff t}\bmn\leftR(\theta^\Ga,t\rightR) \theta^3\\
    &= {\bm{v}}_0\leftR(\theta^\alpha,t\rightR) + {\bm{v}}_1\leftR(\theta^\alpha,t\rightR)\, \theta^3~, \label{eq:velo_eulerian_nocomp}
\end{align}
where 
\begin{equation}
    \bmv_0\leftR(\theta^\Ga,t\rightR) = \frac{\ddiff }{\ddiff t} \bmx_0\leftR(\theta^\Ga,t\rightR) = \hat{\bmv}_0\leftR(\xi^\Gb\leftR(\theta^\Ga,t\rightR)\rightR)~, \label{eq:v0_dx0dt_def} 
\end{equation}
is the mid-plane velocity and 
\begin{equation}
    \bmv_1\leftR(\theta^\Ga,t\rightR) = \frac{\ddiff }{\ddiff t} \bmn\leftR(\theta^\Ga,t\rightR) = \hat{\bmn}\leftR(\xi^\Gb\leftR(\theta^\Ga,t\rightR)\rightR)~,\label{eq:v1_dndt_def} 
\end{equation}
is the first-order velocity.
Equation~\eqref{eq:velo_eulerian_nocomp} can be decomposed further as
\begin{align}
    {\bm{v}}\leftR(\theta^i,t\rightR) &= v_0^\alpha \bm{a}_\alpha + v_0^3 \bm{n} + v_1^\alpha \bm{a}_\alpha \, \theta^3~, \label{eq:velo_eulerian}
\end{align}
with the first-order velocity components given by
\begin{align}
    v_1^\alpha = \bma^\Ga \cdot \dot{\bmn} = -\left(v^{3}_{0,\Gb} + v_0^{\lambda} b_{\lambda\Gb}\right)a^{\Ga\Gb}~,\label{eq:v1_Eulerian}
\end{align}
where we used $\dot{\bmn}\cdot \bmn = 0$ and $\frac{\ddiff }{\ddiff t}\left(\bmn \cdot \bma_\Ga\right) 
= \dot{\bmn}\cdot\bma_\Ga + \bmn\cdot\bmv_{0,\Ga}$ (see \SMsec{1.1} for details).\textspace%

The mid-surface velocity $\bmv_0$ in Eq.~\eqref{eq:v0_dx0dt_def} can be further expressed as,
\begin{align}
    \bm{v}_0\leftR(\theta^\alpha,t\rightR) &= \frac{\ddiff \bmx_0\leftR(\theta^\alpha,t\rightR)}{\ddiff t} = \frac{\partial \bmx_0\leftR( \theta^\alpha, t\rightR)}{\partial t}\biggr\rvert_{\theta^\gamma} +  \frac{\partial \theta^\beta}{\partial t}\biggr\rvert_{\xi^\gamma} \, \bma_\Gb~. \label{eq:v0_dxdt}
\end{align}%
In \SMsec{1.2}, we argue that a surface parametrization should be considered Eulerian if it satisfies the relation
\begin{align}
    v_0^3 \, \bm{n} = \frac{\partial \bmx_0\leftR( \theta^\alpha, t\rightR)}{\partial t}\biggr\rvert_{\theta^\beta}~, \label{eq:v03_param}
\end{align}
which further implies 
\begin{align}
    v_0^\alpha = \frac{\partial \theta^\alpha}{\partial t}\biggr\rvert_{\xi^\mu}~.\label{eq:v0alpha_Eulerian}
\end{align}%
According to Eqs.~\eqref{eq:v0_dxdt}--\eqref{eq:v0alpha_Eulerian}, a point $\bmx_0\leftR(\bar{\theta}^\Ga,t\rightR)$ with fixed coordinates $\bar{\theta}^\Ga$ is displaced by a normal velocity but not convected by an in-plane velocity. This provides the rationale for referring to $\left\{\theta^\Ga\right\}_{\Ga=1,2,}$ as an Eulerian parametrization and makes it suitable to capture the in-plane fluidity of lipid membranes.\textspace

Using the position vectors in Eqs.~\eqref{eq:KL_Lagrange} and~\eqref{eq:KL_Eulerian}, we can now express the co- and contravariant basis vectors in terms of the mid-surface curvature tensor and the basis $\left\{\check{\bmg}_i\right\}_{i=1,2,3}$. In the following, we only discuss the Eulerian parametrization but the form of the position vectors in Eqs.~\eqref{eq:KL_Lagrange} and~\eqref{eq:KL_Eulerian} implies that equivalent expressions hold for the Lagrangian parametrization. The covariant basis vectors are found from Eqs.~\eqref{eq:g_i_def} and~\eqref{eq:KL_Eulerian}, yielding
\begin{align}
    \bm{g}_\alpha &= \bm{a}_\alpha - b_{\alpha}^{\beta} \bm{a}_\beta \theta^3~, \label{eq:g_a}\\
    \bm{g}_3 &= \bm{n}~, \label{eq:g_3}
\end{align}
where we used the Weingarten formula $\bm{n}_{,\alpha} = -b_\alpha^\beta \bm{a}_\beta$\footnote{ From Eq.~\eqref{eq:g_a}, it is now apparent that the basis $\{\check{\bm{g}}_i\}_{i=1,2,3}$ in Eq.~\eqref{eq:checkg_def} can generally not be induced by a parametrization.}.
According to Eq.~\eqref{eq:g_3}, the basis vector $\bmg_3$ is independent of $\theta^3$, implying that the normal vectors to the interfaces $\mathcal{S}^\pm$ are given by $\bar{\bm{n}}^\pm = \pm\bm{n}$. The covariant metric components $g_{ij}$ follow directly from Eqs.~\eqref{eq:g_ij_def},~\eqref{eq:g_a}, and~\eqref{eq:g_3} with $g_{\alpha 3} = g_{3 \alpha} = 0$. This block diagonal structure shows that finding the contravariant metric components requires inverting $\left[g_{\alpha\beta}\right]$, implying that $g^{\alpha\beta}$ and, in consequence, $\bm{g}^\alpha$ are rational polynomials in $\theta^3$. However, rational polynomials are inconvenient to use in the dimension reduction procedure as it requires the evaluation of integrals along $\theta^3$. This problem can be alleviated by making use of the series expansion \cite{naghdi1962foundations,song2016consistent}, derived in \SMsec{2.1},
\begin{align}
    \bmg^\Ga = \sum_{m=0}^\infty \bm{a}^\alpha \cdot {\bmb}^{m}  \left(\theta^3\right)^m~. \label{eq:ga_full_expansion}
\end{align}
Furthermore, from $g_{\alpha 3} = g_{3 \alpha} = 0$ and $g_{33} = 1$, it immediately follows that 
\begin{align}
    \bm{g}^3 = \bm{g}_3 = \bm{n}~, \label{eq:g3_n}
\end{align}
showing that the third basis vector is an invariant quantity. 

\subsection{Kinematic Constraints} \label{sec:constraints}
In Sec.~\ref{sec:kinematics}, \KL kinematics was introduced as a set of assumptions describing the kinematics of thin bodies.  These assumptions restrict the permissible deformations to satisfy the form of the position vector in Eq.~\eqref{eq:KL_Lagrange}. In Refs.~\cite{podio1989exact,podio1990constraint}, it was first realized that such kinematic assumptions are generally not compatible with three-dimensional constitutive models since the constrained contributions to the (rate of) strain would be stress-independent. This incompatibility can be avoided by treating the \KL assumptions as constraints on the permissible deformations. These constraints are then enforced by so-called reactive stresses, which are not constitutively defined but must be determined simultaneously with all other unknowns. While this implies that the reactive stresses are not known a-priori, geometric arguments for the constraint and solution spaces allow us to determine which components of the reactive stress tensor are non-zero  \cite{carlson2004geometrically,rajagopal2005nature}. Under suitable assumptions about the order of magnitude of the reactive stresses, discussed in Sec.~\ref{sec:assumptions}, this is sufficient for the derivation of the dimensionally-reduced theory.\textspace

To recast the \KL assumptions as a set of equivalent constraints, recall that they imply that the covariant metric tensor in the Lagrangian description takes the form
\begin{align}
    \hat{g}_{{i}{j}} &= \hat{\bm{g}}_{{i}} \cdot \hat{\bm{g}}_{{j}} =
    \begin{bmatrix}
    \left[\hat{g}_{\alpha\beta}\right]  & \bm{0} \\
    \bm{0} & 1
    \end{bmatrix}_{ij}~. \label{eq:general_KL_metric}
\end{align}
In \SMsec{3.1}, we also show that a metric tensor of the form in Eq.~\eqref{eq:general_KL_metric} is obtained if and only if the \KL assumptions are satisfied. 
Therefore, assuming \KL kinematics is equivalent to constraining the metric tensor in the Lagrangian parametrization to satisfy
\begin{align}
    \hat{g}_{\alpha 3} = \hat{g}_{3\alpha} &= 0~, \label{eq:g_a3_cond_1}\\
    \hat{g}_{33} = 1~. \label{eq:g_33_cond_1}
\end{align}

In what follows, we first pose Eqs.~\eqref{eq:g_a3_cond_1} and~\eqref{eq:g_33_cond_1} as constraints with respect to deformations from a reference configuration and subsequently proceed to reformulate them in a rate form. To this end, consider a time-independent reference configuration $\mathcal{M}_0 = \left\{\hat{\bmX}\leftR(\xi^i\rightR) : \xi^i \in \hat{\Omega} \times \hat{\Xi}\right\}$ with basis vectors $\hat{\bmG}_j = \frac{\partial \hat{\bmX}\leftR(\xi^i\rightR)}{\partial \xi^j}$ defined analogously to Eq.~\eqref{eq:g_i_def}. An infinitesimal line element in this reference configuration, $\ddiff{\hat{\bmX}}$, is mapped to a line element $\ddiff{\hat{\bmx}}$ in the current configuration $\mathcal{M}$ by the deformation gradient $\hat{\bmF}$ through \cite{chadwick1999continuum}
\begin{align}
    \ddiff{\hat{\bmx}} = \hat{\bmF} \ddiff{\hat{\bmX}}~. \label{eq:dxFdX}
\end{align}
Equation~\eqref{eq:dxFdX} implies that the deformation gradient $\hat{\bmF}$ can be calculated as the gradient of $\hat{\bmx}$ with respect to the reference configuration $\hat{\bmX}$, i.e. $\hat{\bmF} = \frac{\partial \hat{\bmx}}{\partial \hat{\bmX}}\bigr\rvert_t$. By noting that $\hat{\bmx}\leftR(\xi^i,t\rightR)$ and $\hat{\bmX}\leftR(\xi^i,t\rightR)$ are both functions of the Lagrangian parametrization, the chain rule implies that the deformation gradient can also be written as
\begin{align}
    \hat{\bm{F}} = \hat{\bm{g}}_{{i}} \otimes \hat{\bm{G}}^{{i}}~.
\end{align}

Now consider the length of an infinitesimal line element in the current configuration, $\ddiff{s}^2= \ddiff{\hat{\bmx}} \cdot \ddiff{\hat{\bmx}}$. If we write the corresponding line element in the reference configuration as $\ddiff{\hat{\bmX}} = \bml \diff{S}$ with unit vector $\bml$ and length $\ddiff{S}$, we can use Eq.~\eqref{eq:dxFdX} to relate $\ddiff{s}$ to $\ddiff{S}$ \cite{chadwick1999continuum}, i.e.
\begin{align}
    \ddiff{s}^2 = \bml \cdot \hat{\bmC} \bml \diff{S}^2~, \label{eq:ds2_dS2}
\end{align}
where we defined the right Cauchy-Green tensor $\hat{\bm{C}}$ as
\begin{align}
    \hat{\bm{C}} = \hat{\bm{F}}^T \hat{\bm{F}} = \hat{g}_{ij} \, \hat{\bm{G}}^{{i}} \otimes \hat{\bm{G}}^{{j}}~. \label{eq:Cdef}
\end{align}
From Eq.~\eqref{eq:ds2_dS2}, it is apparent that the right Cauchy-Green tensor is related to the stretch of a line element between the current and reference configurations.\textspace

The definition of the right Cauchy-Green tensor in Eq.~\eqref{eq:Cdef} further shows that Eqs.~\eqref{eq:g_a3_cond_1} and~\eqref{eq:g_33_cond_1} are equivalent to
\begin{align}
    \hat{\Gamma}_{\alpha}\big(\hat{\bm{C}}\big) = \hat{\bm{G}}_{\alpha} \cdot \hat{\bm{C}} \hat{\bm{G}}_{3} = \hat{\bm{G}}_{3} \cdot \hat{\bm{C}} \hat{\bm{G}}_{\alpha} &= 0~, \label{eq:constraint_a_C}\\
     \hat{\Gamma}_{3}(\hat{\bm{C}}) = \hat{\bm{G}}_{3} \cdot \hat{\bm{C}} \hat{\bm{G}}_{3} - 1 &= 0~. \label{eq:constraint_3_C} 
\end{align}
Equations~\eqref{eq:constraint_a_C} and~\eqref{eq:constraint_3_C} can be written more compactly in terms of the Green-Lagrange strain $\hat{\bmE}$, defined as 
\begin{align}
    \hat{\bmE} &= \frac{1}{2}\left(\hat{\bmC} - \bm{1}\right) = \frac{1}{2}\left(  \hat{g}_{ij} - \hat{G}_{ij} \right) \hat{\bm{G}}^{{i}} \otimes \hat{\bm{G}}^{{j}}~, \label{eq:E_def}
\end{align}
where $\hat{G}_{ij}$ is the metric tensor associated with the reference configuration. If the reference configuration satisfies the \KL conditions in Eqs.~\eqref{eq:g_a3_cond_1} and~\eqref{eq:g_33_cond_1}, Eqs.~\eqref{eq:constraint_a_C} and~\eqref{eq:constraint_3_C} take the form 
\begin{align}
    \hat{\Gamma}_i\big(\hat{\bmE}\big) = \hat{\bm{G}}_{i} \cdot \hat{\bm{E}} \hat{\bm{G}}_{3} = \hat{\bm{G}}_{3} \cdot \hat{\bm{E}} \hat{\bm{G}}_{i} &= 0~. \label{eq:constraint_i_E}
\end{align}

The constraints in Eq.~\eqref{eq:constraint_i_E} can also be expressed in terms of the velocity gradient. To that end, we note that the right Cauchy-Green tensor ${\bmC}$ is related to the symmetric part of the velocity gradient $\bmD = \frac{1}{2}\left(\nabla \bmv + \left(\nabla \bmv\right)^T\right)$ by \cite{gurtin2010mechanics}
\begin{align}
    2\bm{D} &= \bm{F}^{-T} \dot{\bm{C}} \bm{F}^{-1} \\
    &= \left( {\bm{g}}^i \otimes {\bm{G}}_i \right) \dot{{\bm{C}}} \left( {\bm{G}}_j \otimes {\bm{g}}^j \right)~, \label{eq:D_C_relation}
\end{align}
now written using the Eulerian parametrization. Differentiation of Eq.~\eqref{eq:constraint_i_E} with respect to time thus shows that the constraints imposed by the \KL assumptions can be equivalently expressed as
\begin{align}
    \Gamma_i\leftR(\bm{D}\rightR) \coloneqq \bm{g}_i \cdot \bm{D} \bm{g}_3 = \bm{g}_3 \cdot \bm{D} \bm{g}_i &=0~, \label{eq:constraint_i_D}
\end{align}
as can be verified by substituting Eq.~\eqref{eq:D_C_relation} into Eq.~\eqref{eq:constraint_i_D}.\textspace 

Thus, we find that imposing \KL kinematics is equivalent to constraining either the Green-Lagrange strain according to Eq.~\eqref{eq:constraint_i_E}, or the symmetric part of the velocity gradient according to Eq.~\eqref{eq:constraint_i_D}. Using $\hat{\bmE}$ or $\bmD$ to describe the constraints is motivated by their common use in studies of constrained continua \cite{carlson2004geometrically,rajagopal2005nature}. In this article, we make use of the formulations in terms of both $\hat{\bmE}$ and $\bmD$---the deformation-based formulation in Eq.~\eqref{eq:constraint_i_E} is convenient to understand the order of magnitude of the reactive stresses in Sec.~\ref{sec:assumptions} while the rate form in Eq.~\eqref{eq:constraint_i_D} allows us to incorporate the reactive stresses into the balance laws, which are formulated in terms of the Eulerian description.\textspace

\subsection{Three-Dimensional Mechanics} \label{sec:balance_laws}
This section provides a summary of the electromechanics of three-dimensional continuous bodies. We begin by discussing the mass, linear, and angular momentum balances without specifying the constitutive behavior. Subsequently, we introduce Maxwell stresses, required to capture the electromechanical coupling of lipid membranes. After that, the form of the stresses associated with enforcing the constraints introduced in Sec.~\ref{sec:constraints} is derived. Finally, we describe the interface conditions between the body $\mathcal{M}$ and the bulk domains $\mathcal{B}^\pm$ as well as the boundary conditions on the lateral surface $\mathcal{S}_{||}$.

\paragraph{Mass balance}
In the Eulerian description, the local form of the mass balance for the body $\mathcal{M}$ reads \cite{gurtin2010mechanics}
\begin{align}
    \dot{\rho} + \rho \divv{\bm{v}} = 0~, \quad \forall \bm{x} \in \mathcal{M}~, \label{eq:mass_balance_eulerian}
\end{align}
where $\rho$ is the mass density, $\dot{\rho} = \ddiff \rho/\ddiff t$ denotes the material time derivative of $\rho$, and $\bmv$ is the velocity vector. In the Lagrangian description, the mass balance takes the form \cite{gurtin2010mechanics}
\begin{align}
    \frac{\hat{\rho}_\mathrm{r}}{\hat{\rho}} = \frac{\ddiff \hat{v}}{\ddiff \hat{V}}~, \quad \forall \bm{x} \in \mathcal{M}~, \label{eq:mass_balance_lagrange}
\end{align}
where $\hat{\rho}_\mathrm{r}$ and $\ddiff \hat{V}$ are the density and an infinitesimal volume in some reference configuration, respectively, and $\ddiff \hat{v}$ is the corresponding infinitesimal volume in the current configuration. The in-plane fluidity of lipid membranes motivates the use of the Eulerian description in large parts of this article. Nonetheless, Eq.~\eqref{eq:mass_balance_lagrange} will prove convenient to derive the $(2+\delta)$-dimensional mass balance in Sec.~\ref{sec:red_mass}.

\paragraph{Linear momentum balance}
With $\dot{\bm{v}} = \ddiff \bm{v}/\ddiff t$, and $\bm{\sigma}$ and $\bm{f}$ denoting the Cauchy stress and body force per unit volume, respectively, the linear momentum balance in the Eulerian description reads \cite{gurtin2010mechanics}
\begin{align}
    \rho \dot{\bm{v}} = \divv{\bm{\sigma}^T} + \bm{f}~, \quad \forall \bm{x} \in \mathcal{M}~. \label{eq:3D_lin_mom}
\end{align}
In Eq.~\eqref{eq:3D_lin_mom}, the body force $\bmf$ is not written per unit mass to avoid cumbersome expressions in later sections. However, we could similarly express the body force as $\bmf = \rho \tilde{\bmf}$, where $\tilde{\bmf}$ is the body force per unit mass.

\paragraph{Angular momentum balance}
For the purpose of this article, it is convenient to express the angular momentum balance for non-polar media in the Eulerian description as 
\begin{align}
    \bm{g}_i \times \left( \bm{\sigma}^T \bm{g}^i\right) = \bm{0}~, \quad \forall \bm{x} \in \mathcal{M}~. \label{eq:angular_mom_bal}
\end{align} 
This form of the angular momentum balance is less common and therefore derived in \SMsec{4}. By writing the stress tensor in a generic basis $\left\{\bm{g}_i^\prime\right\}_{i=1,2,3}$ as $\bm{\sigma} = \sigma^{\prime \, ij} \, \bm{g}_i^\prime \otimes \bm{g}^{\prime}_j$, we find that Eq.~\eqref{eq:angular_mom_bal} also implies the usual symmetry condition of the stress tensor \cite{gurtin2010mechanics} (also see \SMsec{5.4})
\begin{align}
    \sigma^{\prime \, ij} = \sigma^{\prime \, ji}~. \label{eq:sigij_sym}
\end{align}
Although the form in Eq.~\eqref{eq:angular_mom_bal} is less widely used, it will prove convenient when applying the dimension reduction procedure to the angular momentum balance in Sec.~\ref{sec:red_angmom}.

\paragraph{Constitutive models}
The equations governing the mechanical behavior, Eqs.~\eqref{eq:mass_balance_eulerian},~\eqref{eq:3D_lin_mom}, and~\eqref{eq:angular_mom_bal}, are only well-posed upon choosing constitutive relations specifying the stress tensor. In this article, we assume the generic relation
\begin{align}
    \bm{\sigma} = \bm{\mathcal{F}}\leftR(\rho, \bm{C}, \bm{D}, \bm{e}, ...\rightR)~, \label{eq:generic_const_model}
\end{align}
where $\bm{e}$ denotes the electric field. The precise form of the constitutive model is not relevant in this article, though we will make further simplifying assumptions about the form of the stress tensor in Sec.~\ref{sec:assumptions}. In part 3 of this series of articles, we choose constitutive models that describe the viscous-elastic material response of lipid membranes.

\paragraph{Maxwell stresses}
To model the electromechanical behavior of lipid membranes, we must account for the stresses resulting from electric fields. In particular, electric fields give rise to so-called Maxwell stresses $\bm{\sigma}_\mathrm{M}$, a continuous analog of the Lorentz force (see Ref.~\cite{kovetz2000electromagnetic} for details). As in part 1 \cite{omar2023ES}, we assume that the body $\mathcal{M}$ can be described as a linear dielectric material that does not carry any free charge in its interior. Under the assumption of electrostatics, the Maxwell stress tensor on $\mathcal{M}$ then takes the form \cite{kovetz2000electromagnetic}
\begin{align}
    \bm{\sigma}_\mathrm{M} = \varepsilon_\mathcal{M}\left(\bm{e} \otimes \bm{e} - \frac{1}{2} \left(\bm{e}\cdot\bm{e}\right) \bm{1}\right)~,  \label{eq:maxwell_stress_membrane}
\end{align}
where $\varepsilon_\mathcal{M}$ is the membrane permittivity. The assumption that the interior of $\mathcal{M}$ does not carry any free charge implies \cite{dorfmann2014nonlinear}
\begin{align}
    \divv{\bm{\sigma}_\mathrm{M}} = \bm{0}~, \quad \forall \bmx \in \mathcal{M}~,
\end{align}
such that the Maxwell stresses do not have to be considered in the linear momentum balance in Eq.~\eqref{eq:3D_lin_mom}. Furthermore, the Maxwell stress tensor is symmetric and therefore satisfies the angular momentum balance in Eqs.~\eqref{eq:angular_mom_bal} and~\eqref{eq:sigij_sym}, implying that it does not have to be considered in the angular momentum balance either. However, the Maxwell stress tensor does enter into the boundary conditions discussed below \cite{fong2020transport}, motivating the definition of the total stress tensor as 
\begin{align}
    \bbar{{\bm{\sigma}}} = \bm{\sigma} + \bm{\sigma}_\mathrm{M}~. \label{eq:combined_sig}
\end{align}

Analogously to Eqs.~\eqref{eq:maxwell_stress_membrane} and~\eqref{eq:combined_sig}, we define the Maxwell stress tensor in the bulk domains, again under the assumptions of electrostatics and linear dielectric materials, as 
\begin{align}
    \bm{\sigma}_\mathrm{M\mathcal{B}} = \varepsilon_\mathcal{B}\left(\bm{e} \otimes \bm{e} - \frac{1}{2} \left(\bm{e}\cdot\bm{e}\right) \bm{1}\right)~,
\end{align}
with $\varepsilon_\mathcal{B}$ denoting the bulk permittivity, and the combined mechanical stresses $\bm{\sigma}_\mathcal{B}$ and Maxwell stresses $\bm{\sigma}_\mathrm{M\mathcal{B}}$ in the bulk as 
\begin{align}
    \bbar{{\bm{\sigma}}}_\mathcal{B} = \bm{\sigma}_\mathcal{B} + \bm{\sigma}_\mathrm{M\mathcal{B}}~. \label{eq:combined_sigB}
\end{align}
Using this notation, the tractions on a boundary with outward pointing normal vector $\bmm$ are given by $\bbar{{\bm{\sigma}}}^T \bmm$ and $ \bbar{{\bm{\sigma}}}^T_\mathcal{B} \bmm$ for the membrane and bulk, respectively.

\paragraph{Reactive stresses}
As discussed in Sec.~\ref{sec:constraints}, the constraints in Eqs.~\eqref{eq:constraint_i_E} and~\eqref{eq:constraint_i_D} are associated with \textit{constraint} or \textit{reactive stresses}, denoted by $\bm{\sigma}_\mathrm{r}$. These reactive stresses are not specified by constitutive models but are determined so as to enforce the constraints. Thus, the reactive stresses can generally not be determined a-priori but need to be solved for along with all other unknowns through the balance laws. However, it is possible to determine the form of the reactive stresses associated with Eqs.~\eqref{eq:constraint_i_E} and~\eqref{eq:constraint_i_D}. Following the geometric arguments about the constraint spaces and associated solution spaces in Refs.~\cite{carlson2004geometrically,rajagopal2005nature},
we obtain the reactive stress tensor required to enforce Eqs.~\eqref{eq:constraint_i_E} and~\eqref{eq:constraint_i_D} in the Lagrangian and Eulerian parametrizations, respectively, as
\begin{align}
    \hat{\bm{\sigma}}_\mathrm{r} &= \sum_{j=1}^3 \hat{\tilde{\Lambda}}_j\leftR(\xi^i,t\rightR) \, \bmF \frac{\partial \hat{\Gamma}_j\big(\hat{\bm{E}}\big)}{\partial \hat{\bm{E}}} \bmF^T~, \label{eq:reactive_stresses_E} \\
    \bm{\sigma}_\mathrm{r} &= \sum_{j=1}^3 \tilde{\Lambda}_j\leftR(\theta^i,t\rightR) \, \frac{\partial \Gamma_j\leftR(\bm{D}\rightR)}{\partial \bm{D}}~, \label{eq:reactive_stresses_D}
\end{align}
where $\hat{\tilde{\Lambda}}_j$ and $\tilde{\Lambda}_j$ are the unknown reactive stress components in the Eulerian and Lagrangian parametrization, respectively. Evaluating the derivatives in Eqs.~\eqref{eq:reactive_stresses_E} and~\eqref{eq:reactive_stresses_D} yields the same reactive stress tensor,
\begin{align}
    \bm{\sigma}_\mathrm{r} &= \frac{1}{2}\sum_{j=1}^3 \tilde{\Lambda}_j\leftR(\theta^i,t\rightR) \, \left( \bm{g}_j \otimes \bm{g}_3 + \bm{g}_3 \otimes \bm{g}_j \right)~, \label{eq:constraint_stresses_D_basis_g}
\end{align}
where we have now omitted the Lagrangian form for brevity. It is worth noting that the reactive stress tensor is symmetric and therefore satisfies the angular momentum balance in Eq.~\eqref{eq:sigij_sym}. Furthermore, since the components $\tilde{\Lambda}_j\leftR(\theta^i,t\rightR)$ vary along the thickness direction, and $\text{span}\leftR(\bm{g}_1, \bm{g}_2\rightR) = \text{span}\leftR(\bm{a}_1,\bm{a}_2\rightR)$ and $\bm{g}_3 = \bm{n}$, Eq.~\eqref{eq:constraint_stresses_D_basis_g} can be expressed, without loss of generality, in the basis $\left\{\check{\bmg}_i\right\}_{i=1,2,3}$, defined in Eq.~\eqref{eq:checkg_def}, as
\begin{align}
    \bm{\sigma}_\mathrm{r} = \frac{1}{2}\sum_{j=1}^3 \Lambda_i\leftR(\theta^i,t\rightR) \, \left( \check{\bm{g}}_j \otimes \bm{n} + \bm{n} \otimes \check{\bm{g}}_j \right)~. \label{eq:constraint_stresses_basis_D_basis_a}
\end{align}
This form of the reactive stresses is more suitable for the dimension reduction procedure and is therefore preferred over Eq.~\eqref{eq:constraint_stresses_D_basis_g}.\textspace 

The reactive stress tensor in Eq.~\eqref{eq:constraint_stresses_basis_D_basis_a} is related to the mechanical stress $\bm{\sigma}$ through the additive decomposition \cite{carlson2004geometrically, rajagopal2005nature}
\begin{align}
    \bm{\sigma} = \bm{\sigma}_\mathrm{a} + \bm{\sigma}_\mathrm{r}~,
\end{align}
where $\bm{\sigma}_\mathrm{a}$ is the active or constitutively-prescribed stress. Furthermore, if the constraints are satisfied, the active and reactive stresses are orthogonal \cite{carlson2004geometrically, rajagopal2005nature}, i.e.
\begin{align}
    \bm{\sigma}_\mathrm{a} : \bm{\sigma}_\mathrm{r} = 0~, \label{eq:active_reactive_orthogonality}
\end{align}
where $:$ indicates a double contraction. The orthogonality relation in Eq.~\eqref{eq:active_reactive_orthogonality} implies that the components $\check{\bm{g}}_i \cdot \bm{\sigma} \bm{n}$ and $\bm{n} \cdot \bm{\sigma} \check{\bm{g}}_i$ of the overall mechanical stress are exclusively determined by the reactive stresses in Eq.~\eqref{eq:constraint_stresses_basis_D_basis_a} and cannot be prescribed by constitutive models, i.e. $\check{\sigma}^{3 i} = \check{\sigma}^{i 3} = \check{\sigma}^{3i}_\mathrm{r} =  \check{\sigma}^{i 3}_\mathrm{r}$. Thus, we have now determined the form of the stress tensor required to enforce \KL kinematics and eliminated the incompatibility of the \KL assumptions with three-dimensional constitutive models.

\paragraph{Boundary conditions}
The membrane $\mathcal{M}$ is embedded into the bulk domains $\mathcal{B}^\pm$ as shown in Fig.~\ref{fig:schematic_setup}. At the interfaces between $\mathcal{M}$ and $\mathcal{B}^\pm$, we impose continuity of velocities and tractions, i.e.
\begin{alignat}{2}
    \bm{v}\rvert_{\mathcal{S}^\pm} &= \bm{v}_{\mathcal{B}}\rvert_{\mathcal{S}^\pm}~, \quad &&\forall \bm{x} \in \mathcal{S}^\pm~, \label{eq:velocity_jump_0}\\[6pt]
    \bbar{\bm{\sigma}}^T\bigr\rvert_{\mathcal{S^\pm}}\bar{\bm{n}}^\pm &= \bbar{\bm{\sigma}}_\mathcal{B}^T\bigr\rvert_{\mathcal{S^\pm}}\bar{\bm{n}}^\pm~, \quad &&\forall \bm{x} \in \mathcal{S}^\pm~, \label{eq:tract_cont_0}
\end{alignat}
where $\bm{v}_{\mathcal{B}}$ denotes the velocities in the bulk domains and $\bar{\bm{n}}^\pm$ are the normal vectors to $\mathcal{S}^\pm$, pointing towards $\mathcal{B}^\pm$ \cite{casey2011derivation} (see Fig.~\ref{fig:schematic_setup}). The no-slip and no-penetration conditions in Eq.~\eqref{eq:velocity_jump_0} could be similarly formulated using displacements rather than velocities. However, motivated by the fluidity of lipid membranes, this article treats velocities as the primary variables. We emphasize here that the traction continuity condition in Eq.~\eqref{eq:tract_cont_0} requires the total stress tensors defined in Eq.~\eqref{eq:combined_sig} and~\eqref{eq:combined_sigB}, thus including Maxwell stresses. Hence, by defining the mechanical tractions acting on $\mathcal{S}^\pm$ from the bulk domains as $\bm{t}^\pm = \bm{\sigma}^T_\mathcal{B}\bigr\rvert_{\mathcal{S^\pm}}\bar{\bm{n}}^\pm$, Eq.~\eqref{eq:tract_cont_0} can be recast in terms of mechanical and Maxwell contributions as
\begin{align}
    \bm{\sigma}^T\bigr\rvert_{\mathcal{S^\pm}}\bar{\bm{n}}^\pm + \bm{\sigma}_{\mathrm{M}}^T\bigr\rvert_{\mathcal{S^\pm}}\bar{\bm{n}}^\pm = \bm{t}^\pm + \bm{\sigma}_{\mathrm{M}\mathcal{B}}^T\bigr\rvert_{\mathcal{S^\pm}}\bar{\bm{n}}^\pm~, \quad \forall \bm{x} \in \mathcal{S}^\pm~, \label{eq:tract_cont_1}
\end{align}
where we used the stress definitions in Eqs.~\eqref{eq:combined_sig} and~\eqref{eq:combined_sigB}.
\textspace

To specify boundary conditions on the lateral surface $\mathcal{S}_{||}$, we split $\mathcal{S}_{||}$ into non-overlapping domains $\mathcal{S}_{||\mathrm{D}}$ and $\mathcal{S}_{||\mathrm{N}}$ with imposed velocities $\bar{\bmv}$ and tractions $\bar{\bmt}$, respectively, implying
\begin{alignat}{2}
    \bm{v} &= \bar{\bm{v}}~, \quad &&\forall \bm{x} \in \mathcal{S}_{||\mathrm{D}}~, \label{eq:Spar_D}\\
    \bbar{\bm{\sigma}}^T \bm{\nu} &= \bar{\bm{t}}~, \quad &&\forall \bm{x} \in \mathcal{S}_{||\mathrm{N}}~,\label{eq:Spar_N}
\end{alignat}
where $\bm{\nu}$ is the outward-pointing normal on $\mathcal{S}_{||}$ as shown in Fig.~\ref{fig:schematic_setup}. The domains $\mathcal{S}_{||\mathrm{D}}$ and $\mathcal{S}_{||\mathrm{N}}$ are chosen so that the same type of boundary condition, i.e. either Eq.~\eqref{eq:Spar_D} or Eq.~\eqref{eq:Spar_N}, is imposed through the thickness direction.
Lastly, note that we do not specify any further boundary conditions in the bulk domains as they are immaterial for the remainder of this article.

\section{Dimension Reduction Method} \label{sec:dimred_method}
Given the three-dimensional problem setup, choice of kinematics, and governing equations presented in Sec.~\ref{sec:3DMechanics_thin_films}, we now describe further details required for the derivation of the $(2+\delta)$-dimensional theory. In Sec.~\ref{sec:dim_red_summary}, the dimension reduction procedure proposed in part 1 \cite{omar2023ES} of this series of articles is summarized, and in Sec.~\ref{sec:expansion}, we present expansions of all unknowns and parameters in terms of Chebyshev polynomials. Finally, Sec.~\ref{sec:assumptions} introduces further assumptions necessary to render the $(2+\delta)$-dimensional theory tractable.

\subsection{Summary of the Dimension Reduction Procedure} \label{sec:dim_red_summary}
To introduce the dimension reduction procedure, we first define the weighted inner product of two functions $f\leftR(\Theta\rightR),g\leftR(\Theta\rightR) \in L_2$, $\Theta \in (-1,1)$, as 
\begin{align}
	\langle f\leftR(\Theta\rightR),g\leftR(\Theta\rightR) \rangle = \frac{1}{\pi} \int_{-1}^1 f\leftR(\Theta\rightR) g\leftR(\Theta\rightR) \, \frac{1}{\sqrt{1-\Theta^2}} \diff{\Theta}~. \label{eq:InnerProd_def}
\end{align}
The inner product in Eq.~\eqref{eq:InnerProd_def} defines the orthogonality relation of Chebyshev polynomials $\ChebT{n}{\Theta}$, $n\in \mathbb{N}_0$, \cite{boyd2001chebyshev}
\begin{align}
    \langle \ChebT{k}{\Theta}, \ChebT{l}{\Theta}\rangle = \beta_k \delta_{kl}~,
\end{align}
where
\begin{align}
    \beta_k = 
    \begin{cases}
    1 \quad &\text{if } k = 0~, \\
    \frac{1}{2} \quad &\text{otherwise}~.
    \end{cases}
\end{align}
Chebyshev polynomials form a complete set of polynomials, allowing us to express a 
 function $f\leftR(\Theta\rightR) \in L_2$ as
\begin{equation}
    f\leftR(\Theta\rightR) = \sum_{k=0}^\infty {f}_k P_k\leftR(\Theta\rightR)~, \label{eq:fprojection}
\end{equation}
where ${f}_k = \frac{1}{\beta_k} \langle f\leftR(\Theta\rightR),\ChebT{k}{\Theta}\rangle$ are the expansion coefficients. \textspace

Let $\bm{\mathcal{L}}\leftR(\bm{w}\leftR(\theta^i,t\rightR); \bm{p}\leftR(\theta^i,t\rightR)\rightR)$ denote a vector-valued and nonlinear differential operator, where $\bm{w}\leftR(\theta^i,t\rightR)$ is some sufficiently smooth vector-valued function defined on $\mathcal{M}$, and $\bm{p}\leftR(\theta^i,t\rightR) = \left\{\bmp_k\leftR(\theta^i,t\rightR)\right\}_{k=1,...,N_\mathrm{p}}$ is a set of vector-valued parameters. Further, let $\bm{u}\leftR(\theta^i,t\rightR)$ be the solution to the differential equation
\begin{align}
    \bm{\mathcal{L}}\leftR(\bm{u}; \bm{p}\rightR) = \bm{0}~, \label{eq:L_diffeq}
\end{align}
where we omitted the explicit dependence on the parametrization and time for clarity. It is convenient to express the solution $\bm{u}$ in terms of the basis $\left\{\check{\bm{g}}_i\right\}_{i=1,2,3}$, defined in Eq.~\eqref{eq:checkg_def}, since it is independent of the thickness direction $\theta^3$, i.e.
\begin{align}
    \bm{u} = \check{u}^i\leftR(\theta^j,t\rightR) \check{\bm{g}}_i\leftR(\theta^\alpha,t\rightR)~.
\end{align}
Using Eq.~\eqref{eq:fprojection}, the components $\check{u}^i_k\leftR(\theta^i,t\rightR)$ can be expanded in terms of Chebyshev polynomials as
\begin{align}
	\check{u}^i = \sum_{k=0}^{\infty} \check{u}^i_k\leftR(\theta^\alpha,t\rightR) \ChebT{k}{\Theta}~, \quad \text{where } \Theta = \left(2/\delta\right) \theta^3 \in \left(-1,1\right)~.\label{eq:solution_expansion_full}
\end{align}
In Eq.~\eqref{eq:solution_expansion_full}, the unknown coefficients $\check{u}_k^i \leftR(\theta^\alpha,t\rightR)$ are only dependent on the in-plane parametrization while the dependence on the thickness direction is entirely contained in the Chebyshev polynomials. To obtain a dimensionally-reduced theory, the series expansions in Eq.~\eqref{eq:solution_expansion_full} are truncated at some finite order $N_i$. This truncation is the crucial approximation required for dimension reduction and should generally be validated a-posteriori. However, for thin bodies, it is often reasonable to assume a field does not vary significantly along the thickness direction. Upon truncating the series expansion in Eq.~\eqref{eq:solution_expansion_full} and inserting it into Eq.~\eqref{eq:L_diffeq}, we can take the inner product of Eq.~\eqref{eq:L_diffeq} with the $l$\textsuperscript{th} Chebyshev polynomial to obtain 
\begin{align}
    \check{\bm{g}}_j\leftR(\theta^\alpha,t\rightR) \cdot \left\langle \bm{\mathcal{L}}\leftR( \sum_{i=1}^3 \left[ \sum_{k=0}^{N_i} \check{u}^i_k\leftR(\theta^\alpha,t\rightR) \, \ChebT{k}{\Theta} \right]\check{\bm{g}}_i\leftR(\theta^\alpha,t\rightR) ; \bm{p}\rightR), \ChebT{l}{\Theta} \right\rangle& = 0~, \nonumber \\
    j &\in \left[1,2,3\right],~ \forall l \in \left[0, N_j\right]~, \label{eq:inner_prod_eqs}
\end{align}
where Einstein's summation convention does not hold. 
Equation~\eqref{eq:inner_prod_eqs} yields $N_i+1$ equations for the $N_i+1$ coefficients of each of the three components of $\bm{u}$. These equations do no longer depend on the thickness direction $\theta^3$ and are thus considered dimensionally-reduced. As in Ref.~\cite{omar2023ES}, we call the equations obtained from Eq.~\eqref{eq:inner_prod_eqs} \textit{$(2+\delta)$--dimensional} as they contain information about the thickness direction without explicitly depending on the thickness coordinate.\textspace

Before proceeding, several remarks are in place. To arrive at Eq.~\eqref{eq:inner_prod_eqs}, we used the basis $\left\{\check{\bm{g}}_i\right\}_{i=1,2,3}$ as it is independent of $\theta^3$. This assumption is not strictly necessary but simplifies the evaluation of the inner products in Eq.~\eqref{eq:inner_prod_eqs}. Furthermore, evaluation of Eq.~\eqref{eq:inner_prod_eqs} requires expanding the parameters $\bmp_i$ in terms of Chebyshev polynomials, which was omitted above for clarity. However, we note it is not necessary to truncate the expansions of the parameters.
For Eq.~\eqref{eq:inner_prod_eqs} to yield a tractable set of equations, the truncation orders $N_i$ must be sufficiently low. Upon accepting these orders of truncation, however, the proposed method yields exact solutions up to that order and does not require additional approximations. However, the validity of the low-order expansions is problem-specific and should generally be validated. Finally, we emphasize that the proposed method is not limited to Chebyshev polynomials but could be formulated using any complete and orthogonal set of polynomials defined on a bounded domain. 

\subsection{Choice of Expansions} \label{sec:expansion}
To apply the dimension reduction procedure discussed in Sec.~\ref{sec:dim_red_summary}, it is necessary to expand all unknowns and parameters of the differential equations under consideration in terms of Chebyshev polynomials. For the balance laws described in Sec.~\ref{sec:balance_laws}, expansions must be specified for the position and velocity vectors, density, stress tensor, and body force. Motivated by the fluidity of lipid membranes, these expansions are written using the Eulerian parametrization but could be similarly formulated using the Lagrangian parametrization. We begin by rewriting the expressions for the position and velocity vectors in Eqs.~\eqref{eq:KL_Eulerian} and~\eqref{eq:velo_eulerian} in terms of Chebyshev polynomials, 
\begin{align}
    \bm{x}\leftR(\theta^i,t\rightR) &= \bm{x}_0\leftR(\theta^\alpha,t\rightR) \ChebT{0}{\Theta} + \frac{\delta}{2} \bm{n}\leftR(\theta^\alpha,t\rightR) \ChebT{1}{\Theta}~, \label{eq:pos_Eulerian_cheb}\\
    \bm{v}\leftR(\theta^i,t\rightR) &= \bm{v}_0\leftR(\theta^\alpha,t\rightR) \ChebT{0}{\Theta} + \frac{\delta}{2} v_1^\alpha\leftR(\theta^\alpha,t\rightR) \bm{a}_\alpha\leftR(\theta^\alpha,t\rightR) \ChebT{1}{\Theta}~, \label{eq:velo_Eulerian_cheb}
\end{align}
where we used $\theta^3 = \frac{\delta}{2}\Theta$, $\Theta \in (-1,1)$, and the Chebyshev polynomial definitions
\begin{align}
    \ChebT{0}{\Theta} &= 1~, \\
    \ChebT{1}{\Theta} &= \Theta~.
\end{align}
Note that in Eqs.~\eqref{eq:pos_Eulerian_cheb} and~\eqref{eq:velo_Eulerian_cheb} all quantities but the Chebyshev polynomials exclusively depend on the in-plane parametrization.\textspace

To find a suitable expansion of the density $\rho$, we first note that $\rho$ is fully determined by the deformations of $\mathcal{M}$. Thus, restricting the permissible deformations to satisfy \KL kinematics also restricts the permissible densities. This implies that the density cannot be truncated independently and that we must consider the generic series expansion
\begin{align}
    \rho\leftR(\theta^i,t\rightR) = \sum_{k=0}^\infty \rho_k\leftR(\theta^\alpha,t\rightR) \ChebT{k}{\Theta}~. \label{eq:density_expansion_eulerian}
\end{align}
We can also arrive at Eq.~\eqref{eq:density_expansion_eulerian} by invoking analogous arguments involving the Eulerian form of the mass balance and the restriction of the velocity vector according to Eq.~\eqref{eq:velo_eulerian}. Thus, the expansion coefficients $\rho_k\leftR(\theta^\alpha,t\rightR)$ can be found using the mass balances in Eqs.~\eqref{eq:mass_balance_eulerian} or~\eqref{eq:mass_balance_lagrange} together with the assumption of \KL kinematics in Eqs.~\eqref{eq:KL_Lagrange} or~\eqref{eq:velo_eulerian}, as will be discussed in detail in Sec.~\ref{sec:red_mass}.\textspace

Similar to the density, the expansion order of the stress tensor cannot be chosen freely. Instead, it is determined by the reactive stresses in Eq.~\eqref{eq:constraint_stresses_basis_D_basis_a} and the constitutive model in Eq.~\eqref{eq:generic_const_model} once expansions of all its arguments are known. Thus, we use a generic series expansion of the form
\begin{align}
    \bm{\sigma}\leftR(\theta^i,t\rightR) &= \sum_{k=0}^{\infty} \check{\sigma}_k^{ij}\leftR(\theta^\alpha,t\rightR) \, \check{\bm{g}}_i\leftR(\theta^\alpha,t\rightR) \otimes \check{\bm{g}}_j\leftR(\theta^\alpha,t\rightR) \ChebT{k}{\Theta} \label{eq:stress_tensor_expansion_bar}~,
\end{align}
where employing the basis $\left\{\check{\bmg}_i\right\}_{i=1,2,3}$ is convenient as it simplifies separating the in-plane and thickness coordinates.\textspace

Finally, we are left with finding an expansion for the body force $\bm{f}$. Since $\bmf$ is considered a parameter, we express it using the generic series expansion
\begin{align}
\bmf\leftR(\theta^i,t\rightR) &= \sum_{k=0}^{\infty} {\bm{f}}_k\leftR(\theta^\alpha,t\rightR) \ChebT{k}{\Theta}~. \label{eq:expansion_bodyf}
\end{align}
In Sec.~\ref{sec:red_linmom}, it will become apparent that this choice is of little consequence for the derivation of the dimensionally-reduced balance laws and that we could indeed choose a finite order of truncation.

\subsection{Order of Magnitude Assumptions} \label{sec:assumptions} 
To make the $(2+\delta)$-dimensional theory tractable, we introduce physically motivated assumptions that allow us to neglect higher-order expansion coefficients based on their order of magnitude\footnote{Note that we do not seek to truncate the series expansions introduced in Sec.~\ref{sec:expansion} a-priori. Such truncation would not be permissible due to the orthogonality of Chebyshev polynomials. Instead, we use the order of magnitude assumptions to neglect small terms in the dimensionally-reduced balance laws and equations of motion in Secs.~\ref{sec:dimred_balances} and~\ref{sec:EOMs}, as these are independent of the $\theta^3$ coordinate.}. The first such assumption is of purely geometrical nature: We assume that the principal curvatures $\kappa_\alpha$ are small compared to the thickness of the thin film, i.e.
\begin{align}
    \left( \delta \kappa_\alpha \right)^2 \ll 1~, \quad \alpha = 1,2~. \label{eq:ka_ll1}
\end{align}
Note that Eq.~\eqref{eq:ka_ll1} was also assumed in part 1 \cite{omar2023ES} to render the $(2+\delta)$-dimensional theory of electrostatics tractable. Equation~\eqref{eq:ka_ll1} further implies that the mean and principal curvatures, defined in Eqs.~\eqref{eq:H_kappa} and~\eqref{eq:K_kappa}, are also small compared to the thickness of the thin film, i.e.
\begin{align}
    \left( \delta H \right)^2 &\ll 1~, \label{eq:H2small}\\
    \delta^2 \lvert K \rvert &\ll 1~. \label{eq:Ksmall}
\end{align}

Equation~\eqref{eq:ka_ll1} also allows us to obtain order of magnitude estimates for contractions of the curvature tensor $\bmb$ with both tensors and vectors. To show this, recall the spectral decomposition of $\bmb$ in Eq.~\eqref{eq:bmb_spectral_decomp} and consider an arbitrary tensor $\bmA = A^\Ga_{\,.\,\Gb} \: \bml_\Ga \otimes \bml^\Gb$. For $m\geq 0$, we then find
\begin{align}
    \left( \delta \bm{b}\right)^m \colon \bm{A} = \sum_{\alpha = 1}^2 \left( \delta \kappa_\alpha\right)^m A_{\,.\,\alpha}^\alpha~, \label{eq:bA_contr_full}
\end{align}
where the \textit{colon} denotes the double contraction. Since $\bmi \: \colon\bm{A} = A^\alpha_{\,.\,\alpha}$, Eq.~\eqref{eq:bA_contr_full} further implies
\begin{align}
    \left( \delta \bm{b}\right)^m \colon \bm{A} = \mathcal{O}\leftR( \left(\delta\kappa\right)^m\rightR) \left( \bmi \: \colon \bm{A} \right)~, \label{eq:bm_b_contr}
\end{align}
where $\kappa = \max_{\alpha} \lvert \kappa_\alpha \rvert$. Using Eq.~\eqref{eq:ka_ll1}, Eq.~\eqref{eq:bm_b_contr} specifically indicates
\begin{align}
    \left( \delta \bm{b}\right)^n \colon \bm{A} \ll \left( \delta \bm{b}\right)^{n-2} \: \colon \bm{A}~, \quad n\geq 2~. \label{eq:b2_b_contr}
\end{align}
Similarly, for the contraction $\bmc\cdot\left(\delta\bmb\right)^m\bma^\Ga$, $m\geq 0$, where $\bmc = c^\Gg \bml_\Gg$ is an arbitrary vector, we find
\begin{align}
    \bmc\cdot\left(\delta\bmb\right)^m\bma^\Ga &= \left(c^\Gg \bml_\Gg\right) \cdot \left(\sum_{\Gb=1}^2 \left(\delta\kappa_\Gb\right)^m \, \bml^\Gb \otimes \bml_\Gb \right) \bma^\Ga =  \sum_{\Ga=1}^2 c^\Gb \left(\delta\kappa_\Gb\right)^m \left(\bml_\Gb \cdot \bma^\Ga\right)~,
\end{align}
which further implies
\begin{align}
    \bmc\cdot\left(\delta\bmb\right)^m\bma^\Ga &= \mathcal{O}\leftR( \left(\delta\kappa\right)^m\rightR)  \, \bmc\cdot\bmi\bma^\Ga~,  \label{eq:cbaa}
\end{align}%
and consequently
\begin{align}
    \bmc\cdot\left(\delta\bmb\right)^n\bma^\Ga &\ll \bmc\cdot\left(\delta\bmb\right)^{n-2}\bma^\Ga~, \quad n \geq 2~. 
\end{align}

In addition to the curvature magnitude assumptions above, we presume there exist in-plane length scales characterizing changes in the velocity, $\ell_\mathrm{v}$, curvature, $\ell_\mathrm{c}$, and stresses, $\ell_\mathrm{s}$. We then assume that the curvature varies over a length scale not much smaller than the length scale over which stresses change, i.e. 
\begin{align}
    \frac{\ell_\mathrm{c}}{\ell_\mathrm{s}} &\not\ll 1~,\label{eq:lcls}
\end{align}
where $\not\ll$ means \textit{not much less than}. In physical problems, the three length scales $\ell_\mathrm{c}$, $\ell_\mathrm{v}$, and $\ell_\mathrm{s}$ are coupled through the equations of motion and are therefore not expected to differ significantly, rendering Eq.~\eqref{eq:lcls} a plausible assumption. In addition, we assume that the curvature and stress length scales satisfy
\begin{align}
    \left( \frac{\delta}{\ell_\mathrm{c}} \right)^2 \ll 1~, \label{eq:ellc_ll1} \\
    \left( \frac{\delta}{\ell_\mathrm{s}} \right)^2 \ll 1~,\label{eq:ells_ll1}
\end{align}
implying that the curvature and stresses vary over length scales larger than the thickness\footnote{In Sec.~\eqref{sec:red_EOM}, it becomes apparent that Eqs.~\eqref{eq:ellc_ll1} and~\eqref{eq:ells_ll1} only play a minor role in simplifying the $(2+\delta)$-dimensional theory and could therefore be easily relaxed.}. Equations~\eqref{eq:ellc_ll1} and~\eqref{eq:ells_ll1} should also be regarded as the limit of the validity of any continuum theory of lipid membranes and therefore do not restrict the applicability of the $(2+\delta)$-dimensional theory further. Also note that Eq.~\eqref{eq:ellc_ll1} was similarly assumed in the derivation of the $(2+\delta)$-dimensional theory of the electrostatics of thin films in part 1\cite{omar2023ES}. Furthermore, we assume that the first-order, in-plane velocity is small compared to the mid-surface, in-plane velocity, i.e.
\begin{align}
    \left(\frac{\delta v_1^\Ga}{v_0^\Gb}\right)^2 \ll 1~. \label{eq:va_1_magnitude}
\end{align}
Considering Eq.~\eqref{eq:v1_Eulerian}, it is apparent that there exist pathological cases where Eq.~\eqref{eq:va_1_magnitude} is not satisfied and thus, it should be verified a-posteriori.\textspace

To obtain an analytically tractable theory, further assumptions about the coefficients of the stress tensor components in Eq.~\eqref{eq:stress_tensor_expansion_bar} are required. First, recall from Eqs.~\eqref{eq:constraint_stresses_basis_D_basis_a}--\eqref{eq:active_reactive_orthogonality} that only the components $\check{\sigma}^{\alpha\beta} = \check{\sigma}^{\alpha\beta}_\mathrm{a}$ can be prescribed by constitutive models and that the remaining components $\check{\sigma}^{3i} = \check{\sigma}^{i3} = \check{\sigma}^{3i}_\mathrm{r} = \check{\sigma}^{i3}_\mathrm{r}$ are determined by the reactive stresses. Now suppose that the stress tensor components $\check{\sigma}^{\alpha\beta}$ are given by a constitutive model that depends on some small, non-dimensional parameter $|\mathcal{A}| < 1$ such that the expansion coefficients in Eq.~\eqref{eq:stress_tensor_expansion_bar} satisfy
\begin{align}
    {\check{\sigma}}^{\alpha \beta}_k &= \mathcal{O}\leftR( \mathcal{A}^{\mathcal{Z}_k} \rightR)~, \label{eq:stress_assumption_1}
\end{align}
for some exponent $\mathcal{Z}_k$ that depends on the order of the coefficient $k$. Examples of such small parameters include the mean curvature $\delta H$ or the (non-dimensional) strain rate. We then assume that the coefficients of the stress tensor components ${\check{\sigma}}^{\alpha \beta}_k$, $k\geq 2$, are not larger than the zeroth- and first-order coefficients, i.e. 
\begin{align}
    \mathcal{Z}_k \geq \mathcal{Z}_0 \quad \text{and} \quad \mathcal{Z}_k \geq \mathcal{Z}_1~, \qquad \forall k \geq 2~. \label{eq:stress_assumption_2}
\end{align}
Importantly, the assumptions in Eq.~\eqref{eq:stress_assumption_2} can be verified a-posteriori by substituting the solution, e.g. the velocity field, into the three-dimensional constitutive models.\textspace

To introduce assumptions for the order of magnitude of the expansion coefficients of the reactive stresses, recall that they assume the value required for the deformations to satisfy \KL kinematics. Since the reactive stresses are not known a-priori but instead must be solved for together with all other unknowns, it is generally not possible to obtain estimates for the order of magnitude of the expansion coefficients of the reactive stresses. In the following, we show that we can mitigate this problem by relaxing the constraints in Eq.~\eqref{eq:constraint_i_E} using the penalty method and considering this relaxed problem as a proxy for the constrained problem. From that, we obtain kinematic conditions that allow us to truncate the reactive stresses arising from the penalty formulation. Subsequently, we assume that the reactive stresses resulting from the penalty method are representative of the true reactive stresses. Details of this approach are discussed in \SMsec{3.2} and only the relevant results are summarized in the following.\textspace

To enforce the constraints imposed by \KL kinematics using the penalty method, it is convenient to consider the Lagrangian parametrization and the form of the constraints specified in Eq.~\eqref{eq:constraint_i_E}. Since the penalty method enforces these constraints only weakly, consider a generic expansion of the position vector in the Lagrangian parametrization, not necessarily satisfying \KL kinematics,
\begin{align}
    \hat{\bmx}^\prime = \sum_{k=0}^\infty \hat{\bmx}^\prime_k\leftR(\xi^\alpha,t\rightR) \ChebT{k}{\Theta}~.\label{eq:xdag_arbitrary}
\end{align}
Assuming the material under consideration is hyperelastic\footnote{For hyperelastic materials, the stress tensor can be expressed as the derivative of a stored energy functional $W\big(\hat{\bmE}\big)$, i.e. $\bm{\sigma} \propto \frac{\partial W\leftR(\hat{\bmE}\rightR)}{\partial \hat{\bmE}}$. Note, however, that this assumption is not valid for lipid membranes.}, one may use the principle of virtual work to solve for the position vector (see Ref.~\cite{ciarlet2021mathematical} for details). In the absence of any constraints, the principle of virtual work gives rise to a minimization problem of the form
\begin{align}
    \hat{\bmx}^\ast = \argmin_{\hat{\bmx}^\prime} \Pi\leftR(\hat{\bmE}\bigl(\hat{\bm{x}}^\prime\bigr)\rightR)~,\label{eq:virtual_work_unconstrained}
\end{align}
where $\Pi$ is the virtual work functional and $\hat{\bmE}$ is the Green-Lagrange strain tensor defined with respect to a stress-free reference configuration $\mathcal{M}_0$ (see Sec.~\ref{sec:constraints}). If the constraints in Eqs.~\eqref{eq:constraint_i_E} are accounted for, and $\mathcal{M}_0$ satisfies the \KL assumptions, Eq.~\eqref{eq:virtual_work_unconstrained} becomes the constrained minimization problem
\begin{align}
    \hat{\bmx}^\dagger = \argmin_{\hat{\bmx}^\prime} {\Pi}\leftR(\hat{\bmE}\leftR(\hat{\bm{x}}^\prime\rightR)\rightR) \quad \text{subject to } \hat{\Gamma}_i\leftR(\hat{\bmE}\leftR(\hat{\bm{x}}^\prime\rightR)\rightR) = 0~,~ i=1,2,3~. \label{eq:virtual_work_constrained}
\end{align}
The constraints in Eq.~\eqref{eq:virtual_work_constrained} can be enforced by the method of Lagrange multipliers where the Lagrange multipliers take the role of the reactive stresses. We now seek to relax these constraints using the penalty method. To that end, we define the Green-Lagrange strain tensor $\hat{\bmE}^{\prime\dagger}$ with the constrained solution $\hat{\bmx}^\dagger$ obtained from Eq.~\eqref{eq:virtual_work_constrained} as the reference configuration and introduce quadratic penalty terms for each of the constraints in Eq.~\eqref{eq:constraint_i_E},
\begin{align}
    \Pi^\mathrm{P}\leftR(\hat{\bmE}^{\prime\dagger}\leftR( \hat{\bm{x}}^\prime\rightR)\rightR) = \frac{1}{2} \sum_{i=1}^3 K_i \int_{\mathcal{M}_0} \hat{\Gamma}_i\leftR(\hat{\bmE}^{\prime\dagger}\leftR(\hat{\bm{x}}^\prime\rightR)\rightR)^2 \diff{V}~, \label{eq:PiP}
\end{align}
where $K_i$ are penalty parameters. In principle, the penalty term in Eq.~\eqref{eq:PiP} could also be written in terms of $\hat{\bmE}$ but the use of $\hat{\bmE}^{\prime\dagger}$ permits the assumption of small deformations, as detailed in \SMsec{3.2}. Using the penalty terms in Eq.~\eqref{eq:PiP}, the constrained minimization problem in Eq.~\eqref{eq:virtual_work_constrained} simplifies to the unconstrained minimization problem
\begin{align}
    \hat{\bmx}^\ddagger = \argmin_{\hat{\bmx}^\prime} \left[ \Pi\leftR(\hat{\bmE}\leftR(\hat{\bm{x}}^\prime\rightR)\rightR) + \Pi^\mathrm{P}\leftR(\hat{\bmE}^{\prime\dagger}\leftR( \hat{\bm{x}}^\prime\rightR)\rightR) \right]~, \label{eq:virtual_work_full_dagger}
\end{align}
By assuming that the difference between the solution to Eq.~\eqref{eq:virtual_work_full_dagger} and the constrained solution $\bmx^\dagger$ is small, the reactive stresses resulting from Eq.~\eqref{eq:virtual_work_full_dagger} take the form
\begin{align}
    \hat{\bm{\sigma}}^\ddagger_{\mathrm{r}i} &= \frac{1}{2} K_i \hat{\Gamma}_i\leftR( \hat{\bm{E}}^{\prime\dagger}\bigl(\bm{x}^\ddagger\bigr)\rightR) \, \left( \hat{\check{\bm{g}}}_i \otimes \hat{\bm{n}} + \hat{\bm{n}} \otimes \hat{\check{\bm{g}}}_i\right)~, \label{eq:sriddagger}
\end{align}
where $\hat{\check{\bm{g}}}_i$ are the basis vectors defined in Eq.~\eqref{eq:checkg_def} written in the Lagrangian parametrization. From this, we find that if the conditions (see \SMsec{3.2} for details)
\begin{align}
    \left( \hat{\bmx}_1^\ddagger - \frac{\delta}{2}\hat{\bmn} \right) \cdot \hat{\bmn} &\gg \hat{\bmx}_l^\ddagger \cdot \hat{\bmn}~, \quad \forall l > 2~,  \label{eq:normal_kinematic_assumption1}\\
    \hat{\bmx}_2^\ddagger \cdot \hat{\bmn} &\gg \hat{\bmx}_l^\ddagger \cdot \hat{\bmn}~, \quad \forall l > 2~, \label{eq:normal_kinematic_assumption2}
\end{align}
and
\begin{align}
    \hat{\bmx}_k^\ddagger \cdot \hat{\bma}_\Ga &\gg \hat{\bmx}_l^\ddagger \cdot \hat{\bma}_\Ga~, \quad k = 1,2~,~\forall l > 2~, \label{eq:inplane_kinematic_assumption2}
\end{align}
are satisfied, the reactive stress tensor components obey
\begin{align}
    \check{\sigma}_k^{i3} = \check{\sigma}_k^{3i} \gg \check{\sigma}_l^{i3} = \check{\sigma}_l^{3i}~, \quad k = 0,1~,~\forall l \geq 2~. \label{eq:OoM_penalty}
\end{align}
Equations~\eqref{eq:normal_kinematic_assumption1}--\eqref{eq:inplane_kinematic_assumption2} require that the expansion coefficients $\bm{x}_1^\dagger$ and $\bm{x}_2^\dagger$ deviate from \KL kinematics the most compared to all other $\bm{x}_i^\dagger$. In that case, Eq.~\eqref{eq:OoM_penalty} implies that the zeroth- and first-order reactive stress tensor components dominate all other components which allows truncating the reactive stresses.\textspace

We can gain further insight into the conditions in Eqs.~\eqref{eq:normal_kinematic_assumption1}--\eqref{eq:inplane_kinematic_assumption2}, by considering the tractions acting on the interfaces $\mathcal{S}^\pm$. In the absence of Maxwell stresses, inserting Eq.~\eqref{eq:sriddagger} into Eq.~\eqref{eq:tract_cont_1} yields
\begin{align}
    \hat{\bmt}^\pm = \pm\frac{1}{2}\sum_{\Ga=1}^2 K_\Ga \hat{\Gamma}_\Ga \rvert_{\mathcal{S}^\pm} \hat{\bma}_\Ga \pm  K_3 \hat{\Gamma}_3 \rvert_{\mathcal{S}^\pm} \hat{\bmn}~, \quad \forall \bm{x} \in \mathcal{S}^\pm~,\label{eq:traction_penalty}
\end{align}
where $\hat{\bmt}^\pm$ are the external tractions on $\mathcal{S}^\pm$ written in the Lagrangian parametrization. According to Eq.~\eqref{eq:traction_penalty}, deviations from \KL kinematics must be supported by external tractions and vice versa. Conversely, Eqs.~\eqref{eq:normal_kinematic_assumption1}--\eqref{eq:inplane_kinematic_assumption2} suggest that external tractions predominantly affect the position vector through $\bm{x}_1^\ddagger$ and $\bm{x}_2^\ddagger$ rather than higher-order displacements from \KL kinematics.\textspace

The above discussion strictly only applies to the penalty method for hyperelastic materials. However, we can extend this result by recalling that if the constraints $\Gamma_i$ are enforced exactly, we can regard the reactive stresses as Lagrange multipliers. Thus, motivated by the equivalence of the Lagrange multiplier and penalty method when $K_i \rightarrow \infty$ \cite{wright1999numerical}, we assume Eq.~\eqref{eq:OoM_penalty} also holds when Eq.~\eqref{eq:constraint_i_E} is enforced by the Lagrange multiplier method rather than the penalty method. Nonetheless, the principle of virtual work only applies to conservative systems. For non-conservative systems, there may exist other variational principles, such as Onsager's variational principle \cite{onsager1931reciprocal,doi2011onsager,arroyo2018onsager}, and the discussion follows similar lines. However, when no variational principle exists, we may introduce \textit{ad-hoc} stresses that enforce the constraints, similar to the use of the penalty method for the incompressible Navier-Stokes equation in the context of the Finite Element Method \cite{hughes1979finite}. We can then use that formulation as a proxy for the exactly constrained model to verify Eq.~\eqref{eq:OoM_penalty} analogously to the discussion above. Thus, the above results serve as a motivation for assuming Eq.~\eqref{eq:OoM_penalty} for the remainder of this article, allowing us to truncate the constrained stresses at first order in Sec.~\ref{sec:red_linmom}.\textspace

In summary, this section introduces several conditions for the applicability of the $(2+\delta)$-dimensional theory. First, we require the curvature of the membrane to be small compared to its thickness (see Eqs.~\eqref{eq:ka_ll1}--\eqref{eq:Ksmall}). Second, we assume there exist characteristic length scales pertaining to changes in curvature, velocity, and stresses along the in-plane directions. The relative magnitudes of the velocity and stress characteristic length scales should satisfy Eq.~\eqref{eq:lcls}, and changes of the curvature and stresses should occur over length scales much larger than the thickness (see Eqs.~\eqref{eq:ellc_ll1} and~\eqref{eq:ells_ll1}). Note, however, that the latter assumption could be omitted without significantly increasing the complexity of the $(2+\delta)$-dimensional theory. The last kinematic assumption, Eq.~\eqref{eq:va_1_magnitude}, states that the first-order, in-plane velocity is smaller than the mid-surface velocity. In addition, we make assumptions about the expansion coefficients of the stress tensor. Specifically, we require that the expansion coefficients of the constitutively determined stresses do not increase in magnitude with increasing polynomial order (see Eqs.~\eqref{eq:stress_assumption_1} and~\eqref{eq:stress_assumption_2}). Furthermore, we assume that the reactive stresses are dominated by their zeroth- and first-order coefficients, which we motivated for the case of hyperelastic materials. Lastly, it is worth noting that all but the last assumption can be verified a-posteriori to ensure applicability of the $(2+\delta)$-dimensional theory\footnote{Verifying the assumption on the reactive stresses requires solving the constrained, three-dimensional problem.}. 

\section{Dimensionally-Reduced Balance Laws} \label{sec:dimred_balances}
In the following sections, we use the dimension reduction procedure summarized in Sec.~\ref{sec:dim_red_summary} to derive the $(2+\delta)$-dimensional balance laws. Our discussion begins with the derivation of the $(2+\delta)$-dimensional mass balance in Sec.~\ref{sec:red_mass}. Subsequently, we present the general form of the stress vectors in Sec.~\ref{sec:stress_vectors} and use these results to obtain the $(2+\delta)$-dimensional linear and angular momentum balances in Secs.~\ref{sec:red_linmom} and~\ref{sec:red_angmom}. Throughout this section, only the key steps are discussed with the algebraic details shown in \SMsec{5}.

\subsection{Mass Balance} \label{sec:red_mass}
To derive the $(2+\delta)$-dimensional mass balance, we make use of both the Eulerian and Lagrangian forms of the three-dimensional mass balance in Eqs.~\eqref{eq:mass_balance_eulerian} and~\eqref{eq:mass_balance_lagrange}, respectively. The Lagrangian form allows us to derive the density expansion coefficients relative to the mid-surface density, which can then be determined using the Eulerian form.\textspace

We begin by assuming that for every point $\hat{\bm{x}}_0\in \mathcal{S}_0$ on the mid-surface, the density $\hat{\rho}$ is constant through the thickness if the mid-surface is locally flat at $\hat{\bm{x}}_0$, i.e. if $\hat{H}=0$ and $\hat{K}=0$\footnote{Note that we still permit density variations along the mid-surface when $\hat{H}=0$ and $\hat{K}=0$.}. This motivates the introduction of a locally flat reference configuration for every point $\hat{\bmx}_0 \in \mathcal{S}_0$ for the Lagrangian form of the mass balance in Eq.~\eqref{eq:mass_balance_lagrange}. With the assumption of constant thickness, Eq.~\eqref{eq:mass_balance_lagrange} can then be written as \cite[\SMsec{5.1}]{chien1944intrinsic,green1950equilibrium,naghdi1962foundations,song2016consistent}
\begin{align}
    \frac{\hat{\rho}_\mathrm{r}}{\hat{\rho}}  = \frac{\ddiff \hat{a}_0}{\ddiff \hat{A}_0} \left( 1 - \delta \hat{H} \Theta + \frac{\delta^2}{4}\hat{K} \Theta^2 \right)~, \label{eq:massbal_lagrange_2}
\end{align}
where $\ddiff \hat{a}_0$ and $\ddiff\hat{A}_0$ are corresponding infinitesimal area elements on the mid-surface of the current and reference configurations, respectively. Since the reference configuration is locally flat with density $\hat{\rho}_\mathrm{r}$ that is homogeneous through the thickness, the mid-surface density is given by 
\begin{align}
    \hat{\rho}_\mathrm{mid} = \hat{\rho}_\mathrm{r} \, \frac{\ddiff \hat{A}_0}{\ddiff \hat{a}_0}~, \label{eq:rhomid_def}
\end{align}
implying that Eq.~\eqref{eq:massbal_lagrange_2} can be written as
\begin{align}
    \hat{\rho} = \hat{\rho}_\mathrm{mid}\left( 1 - \delta \hat{H} \Theta + \frac{\delta^2}{4}\hat{K} \Theta^2 \right)^{-1}~. \label{eq:hatrho_inverse_poly}
\end{align}
Since Eq.~\eqref{eq:hatrho_inverse_poly} is invariant to changes of the parametrization, it can be equivalently written using the Eulerian parametrization as 
\begin{align}
    {\rho} = {\rho}_\mathrm{mid}\left( 1 - \delta {H} \Theta + \frac{\delta^2}{4}{K} \Theta^2 \right)^{-1}~, \label{eq:rho_inverse_poly}
\end{align}
which will be used in the following.\textspace 

Equation~\eqref{eq:rho_inverse_poly} shows that ${\rho}$ is a rational polynomial in $\Theta$, thus requiring a series expansion in terms of Chebyshev polynomials to evaluate the integral inner products arising in the dimension reduction procedure. As derived in \SMsec{5.1}, the coefficients of the density expansion in Eq.~\eqref{eq:density_expansion_eulerian} are then given by
\begin{align}
    {\rho}_i &= {\rho}_\mathrm{mid} \sum_{m=0}^\infty \delta^m \sum_{k=0}^{\lfloor m/2 \rfloor } \frac{\left(-1\right)^k}{4^k} \binom{m-k}{m-2k} {K}^k {H}^{m-2k} c_i \Ga_{mi}~, \label{eq:rho_i}
\end{align}
where 
\begin{align}
    \Ga_{mi} &\coloneqq 
    \begin{cases}
    0, \quad & \text{if } m - i \text{ odd}~, \\
    2^{1-m}\binom{m}{(m-i)/2} \quad & \text{if } m-i \text{ even}~. 
    \end{cases}\label{eq:alpha_mn_def}
\end{align} 
Equation~\eqref{eq:rho_i} determines the expansion coefficients of the density relative to the mid-surface density ${\rho}_\mathrm{mid}$. The mid-surface density can, in turn, be determined by substituting  Eq.~\eqref{eq:density_expansion_eulerian} together with Eq.~\eqref{eq:rho_i} with $i\geq 1$ into the Eulerian form of the mass balance in Eq.~\eqref{eq:mass_balance_eulerian} and taking the inner product with the zeroth Chebyshev polynomial,
\begin{align}
    \left\langle \frac{\diff \rho}{\diff t} + \rho \divv{\bmv}, \ChebT{0}{\Theta} \right\rangle = 0~. \label{eq:Eulerian_mass_zeroth_moment}
\end{align}
As shown in \SMsec{5.1}, invoking Eqs.~\eqref{eq:ka_ll1}--\eqref{eq:Ksmall} and~\eqref{eq:va_1_magnitude}   reduces Eq.~\eqref{eq:Eulerian_mass_zeroth_moment} to the same form as the mass balance for surfaces \cite{scriven1960dynamics,sahu2017irreversible}
\begin{align}
    \frac{\ddiff \rho_\mathrm{s} }{\ddiff t} +  \rho_\mathrm{s} \left( v_{0:\Ga}^\Ga- 2  v H \right)= 0~, \label{eq:mass_zeroth_truncated_final}
\end{align}
where the \textit{colon} indicates the surface covariant derivative and we defined
\begin{align}
    \rho_\mathrm{s} = \rho_0 \delta~. \label{eq:rhos_def}
\end{align}
Note that $\rho_\mathrm{s}$ does not correspond to the mid-surface density $\rho_\mathrm{mid}$ but is proportional to the zeroth-order density coefficient and related to $\rho_\mathrm{mid}$ through Eq.~\eqref{eq:rho_i}. Therefore, given $H$ and $K$, solving Eq.~\eqref{eq:mass_zeroth_truncated_final} for $\rho_0$ fully determines the density $\rho$. Lastly, we note that $\rho_\mathrm{s}$ has units of mass per unit area and can thus be considered as an effective surface density.

\subsection{Stress Vectors} \label{sec:stress_vectors}
The linear and angular momentum balances of thin bodies are commonly expressed using stress vectors rather than stress tensors \cite{naghdi1973theory,steigmann1999fluid,sahu2017irreversible}. Following this convention, we first introduce the concept of stress vectors and subsequently express them in terms of the stress tensor expansion coefficients in Eq.~\eqref{eq:stress_tensor_expansion_bar}. We then use these results in Secs.~\ref{sec:red_linmom} and~\ref{sec:red_angmom} to facilitate the derivation of the $(2+\delta)$-dimensional linear and angular momentum balances.\textspace

Let us define a stress vector $\bm{T}^i$ as
\begin{align}
    \bm{T}^i \coloneqq \bm{\sigma}^T \bm{g}^i~, \label{eq:Ti_def}
\end{align}
implying that the stress tensor $\bm{\sigma}$ can be written as 
\begin{align}
\bm{\sigma} = \bm{T}^i \otimes \bm{g}_i~. \label{eq:sigma_Ti}
\end{align}
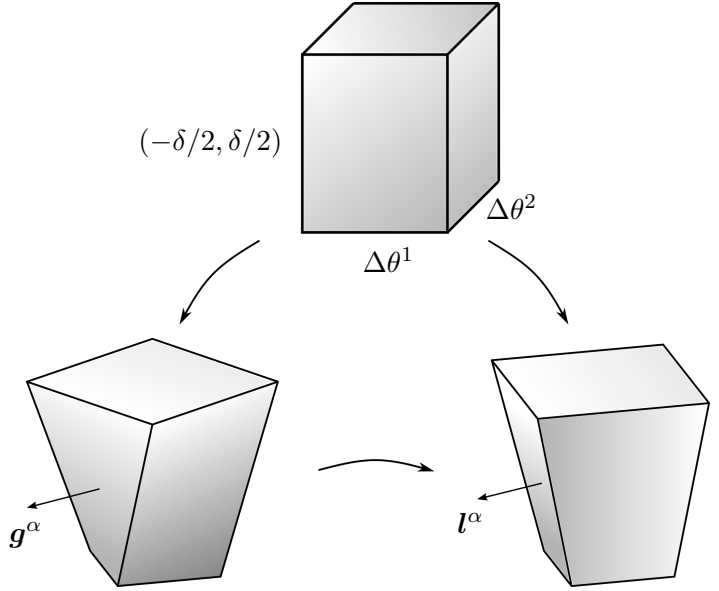
\begin{wrapfigure}{r}{0.55\textwidth}
\centering
\begin{annotationimage}[]{width=1\linewidth}{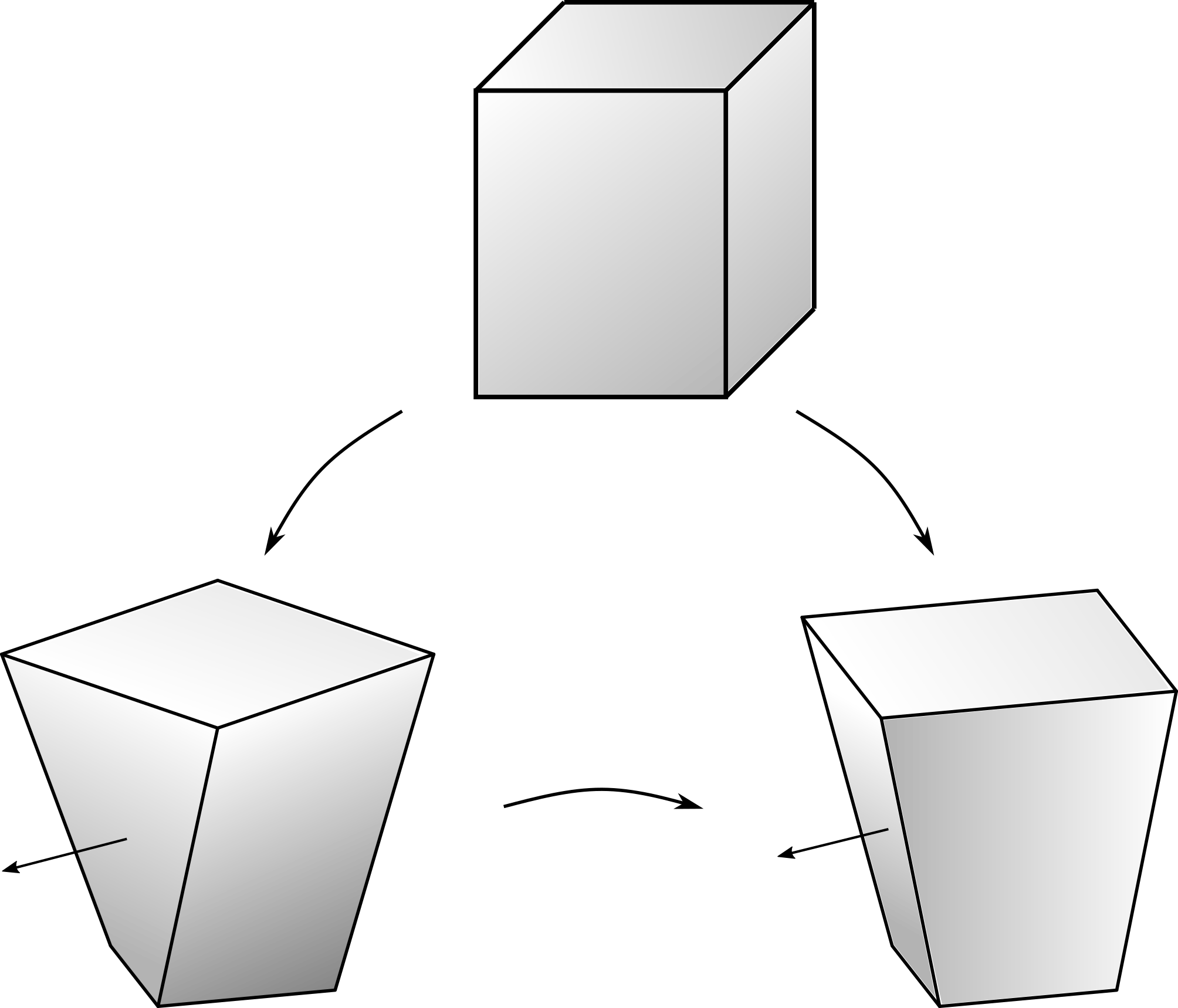}
        \imagelabelset{
                coordinate label style/.style = {
                rectangle,
                fill = none,
                text = black,
                font = \normalfont
        }}
        \draw[coordinate label = {$\bm{g}^\alpha$ at (0.,0.09)}];
        \draw[coordinate label = {$\bm{l}^\alpha$ at (0.65,0.11)}];
        \draw[coordinate label = {$\Delta \theta^1 $ at (0.53,0.56)}];
        \draw[coordinate label = {$\Delta \theta^2 $ at (0.71,0.65)}];
        \draw[coordinate label = {$(-\delta/2,\delta/2)$ at (0.27,0.76)}];
\end{annotationimage}
\caption{A hexahedron in parametric space (top) generally results in a twisted hexahedron in physical space (bottom left). However, when choosing a parametrization such that the basis vectors align with the eigenvectors of the curvature tensor, the resulting image in physical space becomes untwisted (bottom right).}
\label{fig:twisted2cuboid}
\end{wrapfigure}%
Before discussing the expansions of the stress vectors in terms of the stress tensor components, we make a few remarks about the interpretation of stress vectors in curvilinear coordinates. According to Cauchy's stress theorem \cite{chadwick1999continuum}, a stress vector is proportional to the traction acting on an oriented, infinitesimal area element with normal vector $\bm{g}^i/\lvert\lvert \bm{g}^i \rvert\rvert$. If $\bm{g}^i = \bm{e}^i$, where $\left\{\bm{e}^i\right\}_{i=1,2,3}$ is the standard basis of $\mathbb{R}^3$, the associated stress vectors are the tractions acting on an infinitesimal hexahedron \cite{chadwick1999continuum, ciarlet2021mathematical}. In curvilinear coordinates, however, we may consider a hexahedron of the form $\Delta \theta^1 \times \Delta \theta^2 \times (-\delta/2,\delta/2)$ in parametric space (Fig.~\ref{fig:twisted2cuboid} (top)).
The image of this hexahedron in physical space is then twisted, as shown in Fig.~\ref{fig:twisted2cuboid} (bottom left). This follows from the stress vector definition in Eq.~\eqref{eq:Ti_def} and the change of direction of $\bm{g}^\Ga$ along $\theta^3$ according to the basis vector expansion in Eq.~\eqref{eq:ga_full_expansion}.
From Eq.~\eqref{eq:ga_full_expansion}, it is also apparent that the change of direction of $\bmg^\Ga$ arises from the curvature tensor $\bmb$. Hence, we may choose the parametrization of the mid-surface such that the mid-surface contravariant basis vectors coincide with the orthonormal eigenvectors of the curvature tensor $\bml^\Ga$, i.e. $\bma^\Ga = \bml^\Ga$. In this case, the contravariant basis vectors $\bmg^\Ga$ take the form $\bmg^\Ga = \sum_{m=0}^\infty \kappa_\Ga^m \bml^\Ga  \left(\theta^3\right)^m$ (no summation over $\Ga$), implying that $\bmg^\Ga$ does not change its direction but only its magnitude. With this choice of parametrization, the image of the hexahedron in physical space remains untwisted, as shown in the bottom right of Fig.~\ref{fig:twisted2cuboid}. This allows us to interpret the stress vectors in curvilinear coordinates in a manner similar to when using the standard basis.\textspace

Next, we seek to find an expansion of the stress vectors in terms of the stress tensor expansion coefficients. By substituting Eqs.~\eqref{eq:ga_full_expansion} and~\eqref{eq:stress_tensor_expansion_bar} into Eq.~\eqref{eq:Ti_def}, we find that the stress vectors must be expressed using a series expansion, i.e.
\begin{align}
\bm{T}^i = \sum_{n=0}^{\infty} \bm{T}_n^i\leftR(\theta^\alpha\rightR) \ChebT{n}{\Theta}~,\label{eq:stress_vector_expansion}
\end{align}
with expansion coefficients (see \SMsec{5.2} for details)
\begin{align}
    \bm{T}^\alpha_n &=  \check{\bm{g}}_\beta \cdot \frac{1}{2} \left( 
    \sum_{k = 0}^n \check{\sigma}_{n-k}^{\beta j}  \langle  \bm{g}^\alpha, \ChebT{k}{\Theta} \rangle + 
    \sum_{k=0}^\infty  \check{\sigma}_{n+k}^{\beta j}  \langle \bm{g}^\alpha, \ChebT{k}{\Theta} \rangle +
    \sum_{\substack{n> 0 \\ k = n}}^\infty  \check{\sigma}_{k-n}^{\beta j}  \langle  \bm{g}^\alpha, \ChebT{k}{\Theta} \rangle \right) 
    ~ \check{\bm{g}}_j ~, \label{eq:Talpha_n_general}\\[6pt]
    \bm{T}_n^3 &= \check{\sigma}^{3j}_n \check{\bm{g}}_j~. \label{eq:T3_n_general}
\end{align}
Given the order of magnitude assumptions for the stress tensor in Eqs.~\eqref{eq:stress_assumption_1},~\eqref{eq:stress_assumption_2}, and~\eqref{eq:OoM_penalty}, careful inspection of Eq.~\eqref{eq:Talpha_n_general} shows that the order of magnitude of the expansion coefficients $\bm{T}_n^\alpha$, $n\geq 2$, is not larger than the order of magnitude of the stress vectors $\bm{T}_0^\alpha$ and $\bm{T}_1^\alpha$. Furthermore, since the stress vector $\bmT^3$ only depends on the reactive stresses, Eq.~\eqref{eq:OoM_penalty} implies the expansion coefficients $\bm{T}_n^3$, $n\geq 2$, in Eq.~\eqref{eq:T3_n_general} are small compared to $\bm{T}_0^3$ and $\bm{T}_1^3$. Finally, we note that the stress vector definition in Eq.~\eqref{eq:Ti_def} is based on the mechanical stress $\bm{\sigma}$ rather than the total stress $\bbar{\bm{\sigma}}$ in Eq.~\eqref{eq:combined_sig}. This is motivated by the assumption that the body $\mathcal{M}$ does not carry any free charge in its interior, implying that the divergence of the Maxwell stress tensor vanishes in the linear momentum balance (see Sec.~\ref{sec:balance_laws}). However, the above discussion could be extended to account for the effects of Maxwell stresses that are not divergence-free.

\subsection{Linear Momentum Balance} \label{sec:red_linmom}
To derive the $(2+\delta)$-dimensional form of the linear momentum balance, we start from the three-dimensional linear momentum balance in Eq.~\eqref{eq:3D_lin_mom}. For clarity, the inertial, stress divergence, and body force contributions are discussed separately, and the details of the derivations are presented in \SMsec{5.3}.\textspace

\paragraph{Inertia}
To evaluate the inertial term on the left-hand side of Eq.~\eqref{eq:3D_lin_mom}, we consider the velocity and density expansions in Eqs.~\eqref{eq:velo_Eulerian_cheb} and~\eqref{eq:density_expansion_eulerian} together with the density coefficients in Eq.~\eqref{eq:rho_i}. As shown in \SMsec{5.3}, the inertial contribution can then be written to first polynomial order as
\begin{align}
    \delta \rho \bm{a} &\stackrel{1}{\approx}  \rho_{\mathrm{s}} \dot{\bm{v}}_0 \ChebT{0}{\Theta} + \left( \rho_{\mathrm{s}}\delta H \dot{\bm{v}}_0 +  \frac{\delta \rho_{\mathrm{s}}}{2} \,\ddot{\bm{n}} \right) \ChebT{1}{\Theta}~, \label{eq:inertia_reduced01}
\end{align}
where the material time derivatives of the mid-surface velocity and normal vectors are given by 
\begin{align}
    \dot{\bmv}_0& = \left( v^\Ga_{0,t} +  v_0^\Gb v_{0:\Gb}^\Ga - 2v_0^3 v_0^\Gb b_{\Gb}^{\Ga} - v_0^3 v_{0,\Gb}^{3} \, a^{\Ga\Gb} \right) \bma_\Ga + \left( v_{0,t}^3 + 2v_0^\Ga v_{0,\Ga}^3 + v_0^\Ga v_0^\Gb b_{\Gb \Ga} \right) \bmn~, \label{eq:dotv0} \\
    \ddot{\bmn} &= \left( v^\Ga_{1,t} + v_0^\Gb v^\Ga_{1,\Gb} + v_1^\Gb v_{0:\Gb}^\Ga - v_1^\Gb v_0^3 b_\Gb^\Ga -  v_1^\Gb v^\Ga_{0,\Gb} \right) \bma_\Ga + \left( v_1^\Ga v_{0,\Ga}^3 + v_1^\Ga v_0^\Gb b_{\Gb \Ga} \right) \bmn~. \label{eq:ddotn}
\end{align}
In Eq.~\eqref{eq:inertia_reduced01}, the notation $\stackrel{k}{\approx}$ indicates that the expression is truncated after the $k$\textsuperscript{th} polynomial order.
The truncation of Eq.~\eqref{eq:inertia_reduced01} is motivated by Sec.~\ref{sec:red_EOM}, where it will become apparent that we only require the linear momentum balance up to first order.	
When deriving Eq.~\eqref{eq:inertia_reduced01}, we made use of the order of magnitude assumptions in Eqs.~\eqref{eq:H2small},~\eqref{eq:Ksmall},~\eqref{eq:va_1_magnitude}, and~\eqref{eq:rho_i}. This derivation is described in full in \SMsec{5.3}.

\paragraph{Stress divergence}
The stress divergence in Eq.~\eqref{eq:3D_lin_mom} can be analyzed in terms of the stress vectors by using the relation (see \SMsec{5.3}) 
\begin{align}
    \divv{\bm{\sigma}^T} = \bm{T}^i\bigr\rvert_i~, \label{eq:divsig_Tii}
\end{align}
where $\bm{T}^i\bigr\rvert_i$ is the three-dimensional covariant derivative of $\bm{T}^i$ with respect to $\theta^i$. In \SMsec{5.3}, we show that Eq.~\eqref{eq:divsig_Tii} can be written in terms of the Chebyshev expansion coefficients as
\begin{align}
    \bm{T}^i\bigr\rvert_i \stackrel{1}{=}& \, \ChebT{0}{\Theta}\left\{ \bm{T}_{0:\alpha}^\alpha - \frac{1}{4}\sum_{m=0}^\infty \frac{\delta^{m+1}}{2^m} \bma^{\Gb} \cdot \bmb^m\bma_\Gg \, b^\Gg_{\Ga;\Gb} \sum_{k=0}^{m+1} \bmT_k^\Ga \alpha_{(m+1) k} + \frac{2}{\delta} \sum_{k=1}^\infty (2k-1) \bmT_{2k-1}^3 \: - \nonumber \right. \\
    & \left. \hspace{7.8cm} \frac{1}{\delta}\sum_{m=1}^\infty\frac{\delta^{m}}{2^m}  \tr{\leftR(\bmb^{m}\rightR)} \sum_{k=0}^{m-1} \bmT^3_k \alpha_{(m-1) k}\right\} \, + \nonumber \\
    &\, \ChebT{1}{\Theta}\left\{ \bm{T}_{1:\alpha}^\alpha - \frac{1}{4}\sum_{m=0}^\infty \frac{\delta^{m+1}}{2^m} \bma^{\Gb} \cdot \bmb^m\bma_\Gg \, b^\Gg_{\Ga;\Gb} \sum_{k=0}^{m+1}  {c_k} \alpha_{(m+1) k} \left( \bmT^\Ga_{k+1}  + \frac{1}{c_{\lvert k-1\rvert}}\bmT^\Ga_{\lvert k-1 \rvert}\right) \: + \right. \nonumber \\
    & \hspace{1.7cm}\left. \frac{4}{\delta}\sum_{k=1}^\infty 2k \bmT_{2k}^3  -\frac{1}{\delta}\sum_{m=1}^\infty \frac{\delta^{m}}{2^m}  \tr{\leftR(\bmb^{m}\rightR)} \sum_{k=0}^{m-1} c_{k} \alpha_{(m-1) k} \left( \bmT^3_{k+1}  + \frac{1}{c_{\lvert k-1\rvert}}\bmT^3_{\lvert k-1 \rvert} \right) \right\}~,
    \label{eq:stress_divergence_fullexp01}
\end{align}%
where $\Ga_{mk}$ is defined in Eq.~\eqref{eq:alpha_mn_def} and
\begin{align}
    c_k = 
    \begin{cases}
    1/2~, \quad &\text{if } k = 0~,\\
    1~, \quad &\text{otherwise}~.
    \end{cases}
\end{align}
As with the inertial term in Eq.~\eqref{eq:inertia_reduced01}, we only show the stress divergence up to first polynomial order. However, it is worth noting that Eq.~\eqref{eq:stress_divergence_fullexp01} contains stress vector coefficients of all orders. In \SMsec{5.3}, we further show that applying the assumptions in Eqs.~\eqref{eq:H2small}--\eqref{eq:b2_b_contr},~\eqref{eq:lcls},~\eqref{eq:stress_assumption_1},~\eqref{eq:stress_assumption_2}, and~\eqref{eq:OoM_penalty} reduces Eq.~\eqref{eq:stress_divergence_fullexp01} to 
\begin{align}
    \bm{T}^i\bigr\rvert_i \stackrel{1}{\approx} &~ \ChebT{0}{\Theta}\left\{ \bm{T}_{0:\alpha}^\alpha - \frac{\delta}{2}H_{,\Ga} \bmT_1^\Ga + \frac{2}{\delta} \bmT_{1}^3 -2 H \bmT_0^3 \right\} \, + \nonumber \\
    &~ \ChebT{1}{\Theta}\left\{ \bm{T}_{1:\alpha}^\alpha - \delta H_{,\Ga} \left( \bmT_0^\Ga + \frac{1}{2}\bmT_2^\Ga\right) - 2 H \bmT_1^3 - \delta\left(2H^2 - K\right) \bmT_0^3 \right\}~.
    \label{eq:stress_divergence_red1_exp01} 
\end{align}%

\paragraph{Body forces}
Truncating the series expansion of the body force per unit volume in Eq.~\eqref{eq:expansion_bodyf} at first polynomial order simply yields
\begin{align}
    \bm{f} \stackrel{1}{=} \bm{f}_0 \ChebT{0}{\Theta} + \bm{f}_1 \ChebT{1}{\Theta}~. \label{eq:1st_bodyf}
\end{align}

\paragraph{Dimensionally-reduced linear momentum balance}
Combining Eqs.~\eqref{eq:inertia_reduced01},~\eqref{eq:stress_divergence_red1_exp01}, and~\eqref{eq:1st_bodyf}, and taking the inner product with the zeroth- and first-order Chebyshev polynomial of each term yields the following zeroth- and first-order linear momentum balances, respectively,
\begin{align}
    \rho_{\mathrm{s}} \dot{\bm{v}}_0  &= \delta\bm{T}_{0:\alpha}^\alpha 
    -\frac{\delta^2}{2} H_{,\Ga} \bmT_1^\Ga 
    + 2 \bmT_{1}^3 
    - 2\delta H \bmT_0^3 + \bm{f}_{\mathrm{s}}~, \label{eq:0thorder_linmom} \\[6pt]
    {\rho_{\mathrm{s}}\delta H \dot{\bm{v}}_0} +  \frac{\rho_{\mathrm{s}}\delta}{2} \ddot{\bm{n}} &= \delta\bm{T}_{1:\alpha}^\alpha - \delta^2 H_{,\Ga} \left( \bmT_0^\Ga + \frac{1}{2} \bmT_2^\Ga\right) - 2\delta H\bmT_1^3 - \delta^2 \left(2H^2-K\right)\bmT_0^3 + \delta\bm{f}_1~, \label{eq:1storder_linmom}
\end{align}
where we have replaced the approximately equal sign ($\approx$) by a strict equality ($=$), multiplied through by $\delta$, used the definition of $\rho_\mathrm{s}$ in Eq.~\eqref{eq:rhos_def}, and also defined the body force per unit area as
\begin{align}
    \bm{f}_\mathrm{s} = \delta\bm{f}_0~.
\end{align}
The zeroth-order linear momentum balance in Eq.~\eqref{eq:0thorder_linmom} resembles the linear momentum balance of strict surface theories \cite{naghdi1973theory,sahu2017irreversible}. However, a detailed comparison requires finding expressions for the stress vectors $\bmT_0^3$ and $\bmT_1^3$ first and is therefore postponed to Sec.~\ref{sec:red_EOM}.

\subsection{Angular Momentum Balance} \label{sec:red_angmom}

The starting point for the derivation of the $(2+\delta)$-dimensional angular momentum balance is the three-dimensional angular momentum balance in Eq.~\eqref{eq:angular_mom_bal}. By substituting the stress vector definition in Eq.~\eqref{eq:Ti_def} and the stress vector expansion in Eq.~\eqref{eq:stress_vector_expansion} into Eq.~\eqref{eq:angular_mom_bal}, taking the inner product of the normal component with the $n$\textsuperscript{th} Chebyshev polynomial yields the condition
\begin{align}
    \bmT_n^\Ga \cdot \bma^\Gb - \frac{\delta}{4} b^\Ga_\Gg \left( \bmT_{n+1}^\Gg + \bmT_{\lvert n-1 \rvert}^\Gg\right)\cdot \bma^\Gb \quad \text{is symmetric}~. \label{eq:angmom_normal_stressvec}
\end{align}
In \SMsec{5.4}, we show that Eq.~\eqref{eq:angmom_normal_stressvec} is equivalent to the in-plane terms of the symmetry condition in Eq.~\eqref{eq:sigij_sym}, i.e.
\begin{align}
    \check{\sigma}_n^{\Ga\Gb} \quad \text{is symmetric}~. \label{eq:sign_sym}
\end{align}
Therefore, a three-dimensional constitutive model satisfying the three-dimensional angular momentum balance in Eq.~\eqref{eq:angular_mom_bal} also satisfies the normal component of the $(2+\delta)$-dimensional angular momentum balance in Eq.~\eqref{eq:angmom_normal_stressvec}. Consequently, given a suitable constitutive model, Eq.~\eqref{eq:angmom_normal_stressvec} does not have to be considered when solving the $(2+\delta)$-dimensional equations of motion.\textspace

To express Eq.~\eqref{eq:angmom_normal_stressvec} with $n=0$ in the form of the angular momentum balance of strict surface theories \cite{naghdi1973theory,steigmann1999fluid,rangamani2013interaction,sahu2017irreversible}, we define
\begin{align}
    N^{\Ga\Gb} &= \delta \bmT_0^\Ga \cdot \bma^\Gb~, \label{eq:Nab_def}\\
    M^{\Ga\Gb} &= -\frac{\delta^2}{4} \bmT_1^\Ga \cdot \bma^\Gb~. \label{eq:Mab_def}
\end{align}
Here, $N^{\Ga\Gb}$ represents the in-plane components of the $0$\textsuperscript{th}-order stress vector, which is constant through the thickness. On the other hand, $M^{\Ga\Gb}$ represents the in-plane components of the $1$\textsuperscript{st}-order stress vector, which varies linearly through the thickness and thus induces a moment about the mid-surface. With these definitions, Eqs.~\eqref{eq:angmom_normal_stressvec} and~\eqref{eq:sign_sym} with $n=0$ yield
\begin{align}
    \check{\sigma}_0^{\Ga\Gb} = N^{\Ga\Gb} + b^\Ga_\Gg M^{\Gg\Gb} \quad \text{is symmetric}~, \label{eq:direct_normal_angmom1}
\end{align}
or, equivalently, 
\begin{align}
    N^{\Ga\Gb} - b^\Gb_\Gg M^{\Gg\Ga} \quad \text{is symmetric}~. \label{eq:direct_normal_angmom2}
\end{align}
Equation~\eqref{eq:direct_normal_angmom2} shares the same form as the normal component of the angular momentum balance found in Naghdi's general shell theory \cite{naghdi1973theory} and strict surface theories for lipid membranes \cite{steigmann1999fluid,rangamani2013interaction,sahu2017irreversible}. However, we note that the definitions of $N^{\Ga\Gb}$ and $M^{\Ga\Gb}$ differ. \textspace

Next, consider the inner product of the in-plane components of Eq.~\eqref{eq:angular_mom_bal}  with the $n$\textsuperscript{th} Chebyshev polynomial, which yields
\begin{align}
    \bmT_n^\alpha \cdot \bmn -  \frac{\delta}{4}b^{\Ga}_\Gg \left( \bmT_{n+1}^\Gg + \bmT_{\lvert n-1\rvert}^\Gg \right) \cdot \bm{n} - \bmT_n^3 \cdot \bma^\Ga = 0~, \label{eq:inplane_angmom_nmoment}
\end{align}
as derived in detail in \SMsec{5.4}. As above, the symmetry condition in Eq.~\eqref{eq:sigij_sym} implies that Eq.~\eqref{eq:inplane_angmom_nmoment} is equivalent to (see \SMsec{5.4})
\begin{align}
    \check{\sigma}_n^{\alpha 3} \quad \text{is symmetric}~, \label{eq:siga_sym}
\end{align}
which is consistent with the reactive stresses (see Eqs.~\eqref{eq:constraint_stresses_D_basis_g} and~\eqref{eq:active_reactive_orthogonality}).
Equation~\eqref{eq:inplane_angmom_nmoment} can be compared to strict surface theories by defining
\begin{align}
    S^\Ga &= \delta \bmT_0^\alpha \cdot \bmn~, \label{eq:Sa_def}\\
    R^{\Ga} &= -\frac{\delta^2}{4} \bmT_1^\Ga \cdot \bm{n}~. \label{eq:Ra_def}
\end{align}
Using these definitions, Eq.~\eqref{eq:inplane_angmom_nmoment} with $n=0$ becomes
\begin{align}
    S^\Ga + b^\Ga_\Gg R^\Gg - \delta \bmT_0^3 \cdot \bma^\Ga = 0~, \label{eq:inplane_0_angmom}
\end{align}
which seemingly differs from the angular momentum of strict surface theories \cite{naghdi1973theory,steigmann1999fluid,rangamani2013interaction,sahu2017irreversible}\footnote{Note that Naghdi's dimension reduction procedure yields the same form of the angular momentum balance as in Eq.~\eqref{eq:inplane_0_angmom} \cite{naghdi1973theory}.}. However, Eq.~\eqref{eq:inplane_angmom_nmoment} is rewritten in a more familiar form in Sec.~\ref{sec:red_EOM} by employing the first-order linear momentum balance, revealing agreement with strict surface theories in the case of vanishing inertial terms and external tractions. In Sec.~\ref{sec:red_EOM}, we also require the first-order angular momentum balance, which is obtained by evaluating Eq.~\eqref{eq:inplane_angmom_nmoment} with $n=1$, yielding
\begin{align}
    R^\alpha - \frac{\delta^2}{4}b^{\Ga}_\Gg\left(\delta\bm{T}_2^\Gg \cdot \bm{n} + S^\Gg\right) + \frac{\Gd^2}{4} \bm{T}_1^3\cdot \bma^\Ga = 0~. \label{eq:1storder_angmom_red}
\end{align}

\section{$(2+\delta)$-Dimensional Equations of Motion}\label{sec:EOMs}

This section begins with the derivation of the $(2+\delta)$-dimensional boundary conditions on the membrane-bulk interfaces $\mathcal{S}^\pm$ and the lateral surface $\mathcal{S}_{||}$. Subsequently, we obtain the $(2+\delta)$-dimensional equations of motion by combining the linear and angular momentum balances derived in Secs.~\ref{sec:red_linmom} and~\ref{sec:red_angmom}, and the boundary conditions on $\mathcal{S}^\pm$. Throughout this section, we emphasize the differences and similarities between the $(2+\delta)$-dimensional theory and strict surface theories.

\subsection{Boundary Conditions} \label{sec:red_BC}

\paragraph{Membrane-bulk boundary conditions} We first consider the boundary conditions on the interfaces between the body $\mathcal{M}$ and the bulk domains in Eqs.~\eqref{eq:velocity_jump_0} and~\eqref{eq:tract_cont_1}. The velocity continuity condition in Eq.~\eqref{eq:velocity_jump_0} can be expressed in terms of the velocity expansion coefficients by substituting Eq.~\eqref{eq:velo_Eulerian_cheb} and using the relation $\ChebT{k}{\pm 1} = \left(\pm 1\right)^k$, yielding
\begin{alignat}{2}
    v_0^\alpha \pm \frac{\delta}{2} v_1^\alpha &= \bm{v}_{\mathcal{B}}\rvert_{\mathcal{S}^\pm} \cdot \bma^\Ga~, \quad &&\forall \bm{x} \in \mathcal{S}^\pm~, \label{eq:velo_continuity_red_inplane} \\
    v_0^3 &= \bm{v}_{\mathcal{B}}\rvert_{\mathcal{S}^\pm} \cdot \bmn~, \quad &&\forall \bm{x} \in \mathcal{S}^\pm~. \label{eq:velo_continuity_red_oop}
\end{alignat}
Equation~\eqref{eq:velo_continuity_red_inplane} shows that the membrane in-plane velocities generally differ on the interfaces $\mathcal{S}^\pm$. This is a direct result of the membrane velocity varying through the thickness---an effect that cannot be captured in strict surface theories. In contrast, the constant thickness assumption requires the normal velocities to be equal on $\mathcal{S}^\pm$.\textspace

We next show how the traction continuity condition in Eq.~\eqref{eq:tract_cont_1} can be incorporated into the $(2+\delta)$-dimensional balance laws. To that end, we use the stress vector definition in Eq.~\eqref{eq:Ti_def} and their expansions in Eq.~\eqref{eq:stress_vector_expansion} to express the mechanical stress contribution of $\mathcal{M}$ to Eq.~\eqref{eq:tract_cont_1} as
\begin{align}
    \bm{\sigma}^T\bigr\rvert_{\mathcal{S^\pm}}\bar{\bm{n}}^\pm = \sum_{n=0}^\infty \left(\pm 1\right)^{n+1} \bm{T}_n^3~, \label{eq:tract_stressvec_expansion_full}
\end{align}
where we also used $\bar{\bm{n}}^\pm = \pm\bm{n}~$ and $\ChebT{k}{\pm 1} = \left(\pm 1\right)^k$.
With Eq.~\eqref{eq:tract_stressvec_expansion_full} and the order of magnitude assumptions in Eq.~\eqref{eq:OoM_penalty}, Eq.~\eqref{eq:tract_cont_1} simplifies to 
\begin{align}
    \pm \bm{T}_0^3 + \bm{T}_1^3 \pm \bm{\sigma}_\mathrm{M}^T\bigr\rvert_{\mathcal{S}^\pm}\bm{n} = \bm{t}^\pm \pm \bm{\sigma}_{\mathrm{M}\mathcal{B}}^T\bigr\rvert_{\mathcal{S}^\pm}\bm{n}~, \quad \forall \bm{x} \in \mathcal{S}^\pm~, \label{eq:tract_stressvec_expansion_trunc}
\end{align}
which yields the following expressions for the stress vectors,
\begin{align}
    \bm{T}_0^3 &= \frac{1}{2}\left(\bm{t}^+ - \bm{t}^-\right) + \left\langle \left\llbracket \bm{\sigma}^T_\mathrm{M} \right\rrbracket \right\rangle^\mathcal{M} \bm{n}~, \label{eq:T03_traction}\\[4pt]
    \bm{T}_1^3 &= \frac{1}{2}\left(\bm{t}^+ + \bm{t}^-\right) + \frac{1}{2}\left\llbracket \left\llbracket \bm{\sigma}^T_\mathrm{M} \right\rrbracket \right\rrbracket^\mathcal{M} \bm{n}~,\label{eq:T13_traction}
\end{align}
where we defined
\begin{align}
    \left\langle \left\llbracket \bm{\sigma}^T_\mathrm{M} \right\rrbracket \right\rangle^\mathcal{M} &= \frac{1}{2}\left( \left( \bm{\sigma}_{\mathrm{M}\mathcal{B}}\bigr\rvert_{\mathcal{S}^+} - \bm{\sigma}_{\mathrm{M}}\bigr\rvert_{\mathcal{S}^+}\right) + \left(\bm{\sigma}_{\mathrm{M}\mathcal{B}}\bigr\rvert_{\mathcal{S}^-} - \bm{\sigma}_{\mathrm{M}}\bigr\rvert_{\mathcal{S}^-}\right) \right)~, \\
    \left\llbracket \left\llbracket \bm{\sigma}^T_\mathrm{M} \right\rrbracket \right\rrbracket^\mathcal{M} &= \left( \bm{\sigma}_{\mathrm{M}\mathcal{B}}\bigr\rvert_{\mathcal{S}^+} - \bm{\sigma}_{\mathrm{M}}\bigr\rvert_{\mathcal{S}^+}\right) - \left(\bm{\sigma}_{\mathrm{M}\mathcal{B}}\bigr\rvert_{\mathcal{S}^-} - \bm{\sigma}_{\mathrm{M}}\bigr\rvert_{\mathcal{S}^-}\right)~.
\end{align}

It should be noted that the external tractions $\bmt^\pm$ in Eqs.~\eqref{eq:T03_traction} and~\eqref{eq:T13_traction} are usually not given but instead have to be self-consistently solved for through the electromechanical equations describing the bulk domains. Furthermore, since Eqs.~\eqref{eq:T03_traction} and~\eqref{eq:T13_traction} enforce traction continuity, the velocity continuity conditions in Eqs.~\eqref{eq:velo_continuity_red_inplane} and~\eqref{eq:velo_continuity_red_oop} cannot be enforced using the equations of the body $\mathcal{M}$ but they instead serve as boundary conditions for the bulk fluid equations. Finally, recall that according to Eqs.~\eqref{eq:constraint_stresses_basis_D_basis_a} and~\eqref{eq:T3_n_general}, the stress vectors $\bm{T}_0^3$ and $\bm{T}_1^3$ are determined by the reactive stresses rather than by a constitutive model. However, Eqs.~\eqref{eq:T03_traction} and~\eqref{eq:T13_traction} show that $\bm{T}_0^3$ and $\bm{T}_1^3$ can be found through the conditions of traction continuity, thus not requiring the reactive stresses explicitly. In Sec.~\ref{sec:red_EOM}, we further show that we can fully eliminate the reactive stresses from the equations of motion.\textspace

\begin{figure}[t]
     \centering
     \begin{subfigure}[c]{0.4\textwidth}
        \centering
        \imagebox{0.8\textwidth}{\input{electromechanics-article/Figures/meshing_thickness.tikz}}
        \caption{Meshing with explicit thickness.}%
        \label{subfig:meshing_w_thickness}
     \end{subfigure}\hspace{0.1\textwidth}%
     \begin{subfigure}[c]{0.4\textwidth}
         \centering
         \imagebox{0.8\textwidth}{\input{electromechanics-article/Figures/meshing_nothickness.tikz}}
         \caption{Meshing without explicit thickness.}
        \label{subfig:meshing_wo_thickness}
     \end{subfigure}
     \ifcap{\caption{When solving the $(2+\delta)$-dimensional theory numerically, a discretization that explicitly accounts for the finite thickness of the membrane, as shown in (a), may be cumbersome to implement, in particular for moving meshes. Thus, the mesh on the bounding surfaces, $\mathcal{S}^-$ and $\mathcal{S}^+$, can alternatively be collapsed onto the membrane mid-surface $\mathcal{S}_0$, as shown in (b). }}{\vspace{1cm}}
     \label{fig:meshing_thickness}
\end{figure}
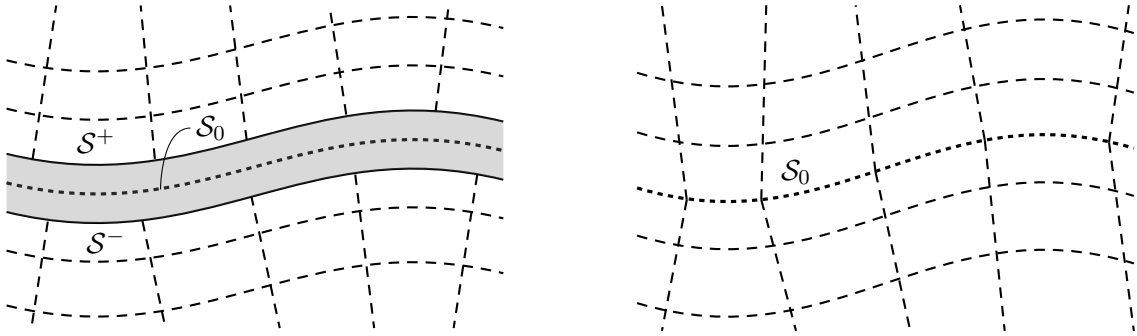

The velocities and tractions in Eqs.~\eqref{eq:velo_continuity_red_inplane},~\eqref{eq:velo_continuity_red_oop}, and~\eqref{eq:tract_stressvec_expansion_trunc} are evaluated on the interfaces $\mathcal{S}^\pm$. However, the balance laws derived in Sec.~\ref{sec:dimred_balances} are independent of the parametric direction $\theta^3$ and can thus be considered as defined on the mid-surface. Thus, when solving the $(2+\delta)$-dimensional balance laws together with the three-dimensional balance laws of the bulk fluid, the body $\mathcal{M}$ can be treated either as a three-dimensional or two-dimensional body. The former implies that bulk quantities are evaluated at points $\bmx_0 \pm \frac{\delta}{2} \bmn$, whereas the latter implies evaluation at $\bmx_0$. These two points of view are illustrated in Fig.~\ref{fig:meshing_thickness}. While the viewpoint in Fig.~\ref{subfig:meshing_w_thickness} is the correct physical interpretation of the $(2+\delta)$-dimensional theory, treating $\mathcal{M}$ as a two-dimensional surface is analytically and numerically simpler. However, this approach introduces additional errors of order $\mathcal{O}\leftR(\delta \kappa\rightR)$, which is of larger order than the order of truncation considered so far. Therefore, treating the membrane as a surface may not be adequate for strongly curved geometries, and the appropriate choice should be made based on the application at hand\footnote{In the examples discussed in part 1 \cite{omar2023ES}, we considered the viewpoint in Fig.~\ref{subfig:meshing_w_thickness}.}.\textspace

\paragraph{Lateral boundary conditions}
We now apply the dimension reduction procedure to the boundary conditions on the lateral surface $\mathcal{S}_{||}$ in Eqs.~\eqref{eq:Spar_D} and~\eqref{eq:Spar_N}. This not only yields boundary conditions that close the equations of motion but also shows that the \KL assumptions restrict the permissible three-dimensional boundary conditions on $\mathcal{S}_{||}$. We begin by taking the inner product of the velocity boundary condition in Eq.~\eqref{eq:Spar_D} with the zeroth-order Chebyshev polynomial, yielding
\begin{alignat}{2}
    v_0^\alpha &= \langle \bar{\bm{v}} \cdot \bm{a}^\alpha, \ChebT{0}{\Theta} \rangle~, \quad &&\forall \bm{x}_0 \in \partial \mathcal{S}_{||\mathrm{D}}~, \label{eq:BCSparD_inplane0}  \\
    v_0^3 &= \langle \bar{\bm{v}} \cdot \bm{n}, \ChebT{0}{\Theta} \rangle~, \quad &&\forall \bm{x}_0 \in \partial \mathcal{S}_{||\mathrm{D}}~, \label{eq:BCSparD_ooplane0}
\end{alignat}
where we defined $\partial\mathcal{S}_{0\mathrm{D}} = \mathcal{S}_{||\mathrm{D}} \cap \partial S_0$.
Equations~\eqref{eq:BCSparD_inplane0} and~\eqref{eq:BCSparD_ooplane0} are Dirichlet boundary conditions prescribing the zeroth-order in-plane and normal velocities, respectively. By taking the inner product of the in-plane components of Eq.~\eqref{eq:Spar_D} with the first-order Chebyshev polynomial, we obtain
\begin{align}
    \frac{\delta}{2} v_1^\alpha &= 2\langle \bar{\bm{v}} \cdot \bm{a}^\alpha, \ChebT{1}{\Theta} \rangle~, \quad \forall\bm{x}_0 \in \partial \mathcal{S}_{||\mathrm{D}}~,
\end{align}
which, according to Eq.~\eqref{eq:v1_Eulerian}, determines the rate of change of the mid-surface normal vector on $\partial \mathcal{S}_{||\mathrm{D}}$. Similarly, taking the inner product of Eq.~\eqref{eq:Spar_D} with higher-order Chebyshev polynomials leads to
\begin{alignat}{2}
    \langle \bar{\bm{v}} \cdot \bm{a}^\alpha, \ChebT{k}{\Theta} \rangle &= 0~, \quad &&\forall k \geq 2,~\forall\bm{x}_0 \in \partial \mathcal{S}_{||\mathrm{D}}~, \label{eq:BCSparD_vak} \\
    \langle \bar{\bm{v}} \cdot \bm{n}, \ChebT{k}{\Theta} \rangle &= 0~, \quad &&\forall k \geq 1,~\forall\bm{x}_0 \in \partial \mathcal{S}_{||\mathrm{D}}~. \label{eq:BCSparD_vk}
\end{alignat}
Equations~\eqref{eq:BCSparD_vak} and~\eqref{eq:BCSparD_vk} do not serve as additional boundary conditions for the velocity $\bmv$ but are necessary conditions for the compatibility of the prescribed velocity $\bar{\bmv}$ with \KL kinematics. Finally, it is worth noting that assigning Eqs.~\eqref{eq:BCSparD_inplane0}--\eqref{eq:BCSparD_vk} the positions $\bmx_0 \in \partial\mathcal{S}_{0\mathrm{D}}$ is rather arbitrary as they hold on the entire domain $\mathcal{S}_{||\mathrm{D}}$. However, this choice is used to indicate that Eqs.~\eqref{eq:BCSparD_inplane0}--\eqref{eq:BCSparD_vk} are independent of the thickness direction and can therefore be regarded as being defined on the boundary of the mid-surface.\textspace

To dimensionally reduce the traction boundary condition on $\mathcal{S}_{||}$ in Eq.~\eqref{eq:Spar_N}, first note that the form of the position vector in Eq.~\eqref{eq:KL_Eulerian} implies that the outward-pointing normal $\bm{\nu}$ is parallel to the mid-surface, i.e. $\bm{\nu} \in \text{span}\leftR(\bm{a}_1, \bm{a}_2\rightR)$, and that $\text{span}\leftR(\bm{g}_1, \bm{g}_2\rightR) = \text{span}\leftR(\bm{a}_1,\bm{a}_2\rightR)$. Therefore, $\bm{\nu}$ can be expressed in either of the following forms
\begin{align}
    \bm{\nu} &= \nu_\Ga\leftR(\theta^i,t\rightR) \bmg^\Ga\leftR(\theta^i,t\rightR) \label{eq:nu_g}\\
    &= \check{\nu}_\Ga\leftR(\theta^\Ga,t\rightR) \bma^\Ga\leftR(\theta^\Ga,t\rightR)~, \label{eq:nu_a}
\end{align}
where the components $\nu_\Ga\leftR(\theta^i,t\rightR)$ and $\check{\nu}_\Ga\leftR(\theta^\Ga,t\rightR)$ are related by 
\begin{align}
    \nu_\Ga = \check{\nu}_\alpha \ChebT{0}{\Theta} - \frac{\delta}{2} \check{\nu}_\beta b_\alpha^\beta \ChebT{1}{\Theta}~.
\end{align}
The form of $\bm{\nu}$ in Eq.~\eqref{eq:nu_g} is more convenient to use for the in-plane components of Eq.~\eqref{eq:Spar_N} while Eq.~\eqref{eq:nu_a} simplifies deriving an expression for the out-of-plane components of Eq.~\eqref{eq:Spar_N}.\textspace

With Eqs.~\eqref{eq:combined_sigB},~\eqref{eq:Ti_def}, and~\eqref{eq:nu_g}, the in-plane components of Eq.~\eqref{eq:Spar_N} can be written as 
\begin{align}
    \bma^\Gb \cdot \bm{T}^\alpha \nu_\alpha = \bma^\Gb \cdot \tilde{\bar{\bm{t}}}~, \quad \forall \bm{x} \in \mathcal{S}_{||\mathrm{N}}~, \label{eq:SparN_deriv_1}
\end{align}
where we defined 
\begin{align}
    \tilde{\bar{\bm{t}}} = \bar{\bm{t}} - \bm{\sigma}^T_{\mathrm{M}}\bm{\nu}~. \label{eq:t_tildebar}
\end{align}
Taking the inner product of Eq.~\eqref{eq:SparN_deriv_1} with the Chebyshev polynomials and using Eqs.~\eqref{eq:Nab_def} and~\eqref{eq:Mab_def} as well as the stress magnitude assumptions in Eqs.~\eqref{eq:stress_assumption_1} and~\eqref{eq:stress_assumption_2} results in the $(2+\delta)$-dimensional boundary conditions 
\begin{alignat}{2} 
    \check{\nu}_\alpha \left( N^{\alpha\beta} + b_\gamma^\alpha M^{\gamma\beta}\right) &= \delta\langle \bma^\Gb \cdot \tilde{\bar{\bm{t}}}, \ChebT{0}{\Theta}\rangle~, \quad &&\forall \bm{x}_0 \in \partial \mathcal{S}_{0\mathrm{N}}~, \label{eq:t0_inplane_BC}\\
    \check{\nu}_\alpha \left( M^{\alpha\beta} + \frac{\delta^2}{16}b^\alpha_\gamma \left(2N^{\gamma\beta} + \delta\bmT_2^\Ga \cdot \bma^\Gb \right)\right) &= -\frac{\delta^2}{2}\langle \bma^\Gb \cdot \tilde{\bar{\bm{t}}}, \ChebT{1}{\Theta}\rangle~, \qquad &&\forall \bm{x}_0 \in \partial \mathcal{S}_{0\mathrm{N}}~, \label{eq:t1_inplane_BC}\\
    \check{\nu}_\alpha \left( \bm{T}_k^\alpha - \frac{\delta}{4} b^\alpha_\beta \left( \bm{T}_{k+1}^\beta + \bm{T}_{k-1}^\beta \right) \right) \cdot \bma^\Gb &= 2\langle \bma^\Gb \cdot \tilde{\bar{\bm{t}}}, \ChebT{k}{\Theta}\rangle~, \quad \forall k \geq 2,~&&\forall \bm{x}_0 \in \partial \mathcal{S}_{0\mathrm{N}}~, \label{eq:tk_inplane_BC}
\end{alignat}
where we defined $\partial \mathcal{S}_{0\mathrm{N}} = \mathcal{S}_{||\mathrm{N}} \cap \partial \mathcal{S}_0$. Now recall that the stress vectors depend on the (rate of) deformation through the choice of constitutive models (see Eq.~\eqref{eq:generic_const_model}). However, the number of fundamental unknowns available under the assumption of \KL kinematics is not sufficient to satisfy Eq.~\eqref{eq:tk_inplane_BC} for arbitrary tractions. Therefore, similarly to Eqs.~\eqref{eq:BCSparD_vak} and~\eqref{eq:BCSparD_vk},  Eq.~\eqref{eq:tk_inplane_BC} should be regarded as a condition on the permissible boundary tractions.\textspace

The normal component of the lateral traction boundary condition in Eq.~\eqref{eq:Spar_N} can be expressed in terms of the tractions $\bmt^\pm$. To see this, we use Eqs.~\eqref{eq:angular_mom_bal},~\eqref{eq:Ti_def}, and~\eqref{eq:nu_a} to reduce the mechanical contribution to the left-hand side of Eq.~\eqref{eq:Spar_N} to
\begin{align}
     \bmn \cdot \left(\bm{\sigma}^T \check{\nu}_\Ga \bma^\Ga\right) = \check{\nu}_\Ga \left( \bm{\sigma}^T \bmn \right) \cdot \bma^\Ga = \check{\nu}_\Ga\, \bmT^3 \cdot \bma^\Ga~.\label{eq:Spar_N_normal}
\end{align}
With this result, taking the inner product of the normal component of Eq.~\eqref{eq:Spar_N_normal} with the Chebyshev polynomials yields
\begin{alignat}{2}
     \frac{\check{\nu}_\alpha}{2}\left( \bm{t}^+ - \bm{t}^-\right)\cdot\bm{a}^\alpha &= \langle \bmn\cdot\tilde{\bar{\bm{t}}}, \ChebT{0}{\Theta}\rangle~, \quad &&\forall \bm{x} \in \mathcal{S}_{||\mathrm{N}}~, \label{eq:t0_oop_BC}\\
    \frac{\check{\nu}_\alpha}{2} \left( \bm{t}^+ + \bm{t}^-\right)\cdot\bm{a}^\alpha &= 2 \langle \bmn \cdot \tilde{\bar{\bm{t}}}, \ChebT{1}{\Theta}\rangle~, \quad &&\forall \bm{x} \in \mathcal{S}_{||\mathrm{N}}~, \label{eq:t1_oop_BC}\\
    \check{\nu}_\Ga \bmT_k^3 \cdot \bma^\Ga &= 2\langle \bmn \cdot \tilde{\bar{\bm{t}}}, \ChebT{k}{\Theta}\rangle~, \quad \forall k>1,~&&\forall \bm{x} \in \mathcal{S}_{||\mathrm{N}}~, \label{eq:tk_oop_BC}
\end{alignat}
where we have used the traction continuity conditions in Eqs.~\eqref{eq:T03_traction} and~\eqref{eq:T13_traction}, and the definition of $\tilde{\bar{\bm{t}}}$ in Eq.~\eqref{eq:t_tildebar}. Now recall that according to Eq.~\eqref{eq:T3_n_general}, the stress vector coefficients $\bmT_k^3$ in Eq.~\eqref{eq:tk_oop_BC} exclusively depend on the reactive stresses. Thus, Eq.~\eqref{eq:tk_oop_BC} can be satisfied by choosing the reactive stresses accordingly, implying that it does not restrict the permissible applied tractions up to the order of magnitude assumptions in Eq.~\eqref{eq:OoM_penalty}.\textspace

From Eqs.~\eqref{eq:t0_inplane_BC} and~\eqref{eq:t1_inplane_BC}, it is now apparent that the stresses $N^{\alpha\beta}$ and moments $M^{\alpha\beta}$ cannot be prescribed independently. Generally, the same can be said in classical thin shell theories but Eqs.~\eqref{eq:t0_inplane_BC} and~\eqref{eq:t1_inplane_BC} differ significantly from the known traction and moment boundary conditions of strict surface theories \cite{green1966further,naghdi1973theory,rangamani2013interaction,sahu2017irreversible}. Similarly, the tractions on $\mathcal{S}_{||\mathrm{N}}$ (i.e. $\tilde{\bar{\bmt}}$) and the tractions on the intersection of $\mathcal{S}_{||\mathrm{N}}$ and $\mathcal{S}^\pm$ (i.e. $\bmt^\pm$) 
must be balanced according to Eqs.~\eqref{eq:t0_oop_BC} and~\eqref{eq:t1_oop_BC} and cannot be prescribed independently. Furthermore, consistency with the three-dimensional theory requires that either Eqs.~\eqref{eq:BCSparD_inplane0}--\eqref{eq:BCSparD_vk} or Eqs.~\eqref{eq:t0_inplane_BC},~\eqref{eq:t1_inplane_BC},~\eqref{eq:t0_oop_BC}, and~\eqref{eq:t1_oop_BC} are enforced but not a combination thereof. This is unlike classical thin shell theory, where, for example, moments and velocities can be prescribed at the same point \cite{naghdi1973theory}. Finally, note that the tractions on the right-hand sides of Eqs.~\eqref{eq:t0_inplane_BC}--\eqref{eq:tk_inplane_BC} and~\eqref{eq:t0_oop_BC}--\eqref{eq:tk_oop_BC} depend on the Maxwell stresses on an edge-like boundary. However, we anticipate that the $(2+\delta)$-dimensional theory cannot accurately resolve Maxwell stresses on such boundaries. In addition, including Maxwell stresses increases the difficulty of rendering the boundary conditions in Eqs.~\eqref{eq:t0_inplane_BC}--\eqref{eq:tk_inplane_BC} compatible with \KL kinematics. Therefore, considering either velocity boundary conditions as in Eqs.~\eqref{eq:BCSparD_inplane0}--\eqref{eq:BCSparD_vk} or closed geometries may often simplify solving the $(2+\delta)$-dimensional theory.

\subsection{Equations of Motion} \label{sec:red_EOM}
Motivated by their close resemblance, we begin this section by using the results obtained in Sec.~\ref{sec:red_BC} to compare the zeroth-order linear momentum balance to the linear momentum balance of strict surface theories. Subsequently, we combine the $(2+\delta)$-dimensional zeroth-order angular momentum balance, and zeroth- and first-order linear momentum balances to obtain the $(2+\delta)$-dimensional equations of motion.\textspace

To compare the zeroth-order linear momentum balance to the linear momentum balance of strict surface theories \cite{naghdi1973theory,rangamani2013interaction,sahu2017irreversible}, we first substitute the expressions for $\bmT_0^3$ and $\bmT_1^3$ in Eqs.~\eqref{eq:T03_traction} and~\eqref{eq:T13_traction} into the zeroth-order linear momentum balance in Eq.~\eqref{eq:0thorder_linmom} to obtain
\begin{align}
    \rho_{\mathrm{s}} \dot{\bm{v}}_0  &= \delta\bm{T}_{0:\alpha}^\alpha 
    -\frac{\delta^2}{2} H_{,\Ga} \bmT_1^\Ga 
    + \left(\bm{t}^+ + \bm{t}^-\right) + \left\llbracket \left\llbracket \bm{\sigma}^T_\mathrm{M} \right\rrbracket \right\rrbracket^\mathcal{M} \bm{n} \nonumber \\
    &\hspace{5cm}
    - 2\delta H \left(\frac{1}{2}\left(\bm{t}^+ - \bm{t}^-\right) + \left\langle \left\llbracket \bm{\sigma}^T_\mathrm{M} \right\rrbracket \right\rangle^\mathcal{M} \bm{n}\right) + \bm{f}_{\mathrm{s}}~. \label{eq:0linmom_traction}
\end{align}
The inertial term on the left-hand side of Eq.~\eqref{eq:0linmom_traction} agrees exactly with that of strict surface theories up to the differing interpretation of the density $\rho_\mathrm{s}$ discussed in Sec.~\ref{sec:red_mass}. Similarly, the stress divergence $\delta\bm{T}_{0:\alpha}^\alpha$ parallels strict surface theories with surface stress vectors replaced by zeroth-order stress vectors. In contrast, the term $-\frac{\delta^2}{2} H_{,\Ga} \bmT_1^\Ga$ is absent in strict surface theories, and its meaning will become more apparent in the derivation of the $(2+\delta)$-dimensional equations of motion below. The external traction and body force terms $\bm{t}^+ + \bm{t}^-$ and $\bmf_\mathrm{s}$ appear analogously in strict surface theories but are usually written as a compound body force \cite{naghdi1973theory,sahu2017irreversible}. 
In contrast, the term $\delta H\left(\bm{t}^+ - \bm{t}^-\right)$ arises from the change of surface area of $\mathcal{S}^\pm$ under bending deformations and is therefore not found in strict surface theories. This new term can be understood by recalling that tractions are forces per unit area. Thus, an increase/decrease in the surface area of $\mathcal{S}^\pm$ arising from curvature changes leads to an increased/decreased force on the stretched/compressed surface. This change of area is directly related to the finite thickness of the body $\mathcal{M}$ and can therefore not be captured in strict surface theories. Finally, as discussed in the introduction in Sec.~\ref{sec:intro}, strict surface theories can generally not describe electromechanical effects and consequently do not capture the terms of Eq.~\eqref{eq:0linmom_traction} associated with Maxwell stresses.\textspace 

Before proceeding with deriving the $(2+\delta)$-dimensional equations of motion, it is worth recalling the fundamental unknowns of our theory. Equation~\eqref{eq:rho_i} shows that the expansion coefficients of the density can be expressed in terms of the mid-surface density and curvature. Thus, it is sufficient to solve for only one of the expansion coefficients, allowing us to consider the density $\rho_\mathrm{s}$, defined in Eq.~\eqref{eq:rhos_def}, as one of our fundamental unknowns. The stress tensor components in Eq.~\eqref{eq:stress_tensor_expansion_bar} are either defined by the constitutive model in Eq.~\eqref{eq:generic_const_model}, or the reactive stresses in Eq.~\eqref{eq:constraint_stresses_D_basis_g} and are thus not fundamental unknowns. Specifically, the in-plane stress tensor components $\check{\sigma}^{\alpha\beta}$ are determined by the constitutive model, which in turn depends on the strain tensor or velocity gradient. According to the assumption of \KL kinematics, the strain tensor and velocity gradient are fully determined by the mid-surface position vector $\bmx_0$ or, equivalently, the mid-surface velocity vector $\bmv_0$. Motivated by the fluid nature of lipid membranes, we consider $\bmv_0$ as a fundamental unknown and determine $\bmx_0$ by integrating Eq.~\eqref{eq:v0_dxdt}. In contrast to the in-plane stress tensor components, the components $\check{\sigma}^{i3} = \check{\sigma}^{3i}$ are determined by the reactive stresses. While we can eliminate some of their occurrences in the $(2+\delta)$-dimensional balance laws by the application of Eqs.~\eqref{eq:T03_traction} and~\eqref{eq:T13_traction}, the stress vector components $\bmT_k^\Ga \cdot \bmn$ also depend on the reactive stresses according to Eq.~\eqref{eq:Talpha_n_general}. However, in the following we show that we can eliminate the reactive stresses from the equations of motions entirely, thereby avoiding explicitly solving for them.\textspace

To derive the equations of motion, we first decompose the zeroth-order linear momentum balance in Eq.~\eqref{eq:0linmom_traction} into its in-plane and out-of-plane components and use the definitions of $N^{\Ga\Gb}$, $M^{\Ga\Gb}$, $S^\Ga$, and $R^\Ga$ in Eqs.~\eqref{eq:Nab_def},~\eqref{eq:Mab_def},~\eqref{eq:Sa_def}, and~\eqref{eq:Ra_def}, respectively, to obtain
\begin{align}
    \rho_\mathrm{s}\left( v^\Ga_{0,t} +  v_0^\Gb v_{0:\Gb}^\Ga - 2v_0^3 v_0^\Gb b_{\Gb}^{\Ga} - v_0^3 v_{0,\Gb}^{3} \, a^{\Ga\Gb} \right) &= N^{\gamma \alpha}_{;\gamma} - S^\gamma b_\gamma^\alpha + 2H_{,\Gg} M^{\Gg\Ga}
    + \left(\bm{t}^+ + \bm{t}^-\right)\cdot \bm{a}^\Ga \nonumber \\ 
    &\hspace{-3.8cm} + \bm{a}^\Ga \cdot \left\llbracket \left\llbracket \bm{\sigma}^T_\mathrm{M} \right\rrbracket \right\rrbracket^\mathcal{M} \bm{n} 
    - 2\delta H \left( \frac{1}{2}\left(\bm{t}^+ - \bm{t}^-\right) + \left\langle \left\llbracket \bm{\sigma}^T_\mathrm{M} \right\rrbracket \right\rangle^\mathcal{M} \bm{n}\right)\cdot\bm{a}^\Ga + \bm{f}_\mathrm{s} \cdot \bm{a}^\Ga~, \label{eq:inplane_eq_1}\\[6pt]
    \rho_\mathrm{s}\left( v_{0,t}^3 + 2v_0^\Ga v_{0,\Ga}^3 + v_0^\Ga v_0^\Gb b_{\Gb \Ga} \right) &= N^{\alpha\beta}b_{\alpha\beta} + S^\alpha_{:\alpha}
    + \left(\bm{t}^+ + \bm{t}^-\right)\cdot \bm{n}  \nonumber \\
    & \hspace{-3.2cm} + \bm{n} \cdot \left\llbracket \left\llbracket \bm{\sigma}^T_\mathrm{M} \right\rrbracket \right\rrbracket^\mathcal{M} \bm{n}
    - 2\delta H \left(\frac{1}{2}\left(\bm{t}^+ - \bm{t}^-\right) + \left\langle \left\llbracket \bm{\sigma}^T_\mathrm{M} \right\rrbracket \right\rangle^\mathcal{M} \bm{n}\right)\cdot\bm{n} + \bm{f}_{\mathrm{s}}\cdot \bmn~. \label{eq:shape_eq_1}
\end{align}
Note that in Eq.~\eqref{eq:shape_eq_1}, we have neglected the term $2 H_{,\Ga} R^\Ga$ as it is small according to Eq.~\eqref{eq:1storder_angmom_red}. Furthermore, the term $-\frac{\delta^2}{2} H_{,\Ga} \bmT_1^\Ga\cdot\bma^\Gb = 2H_{,\Gg} M^{\Gg\Ga}$ in Eq.~\eqref{eq:inplane_eq_1}, absent in strict surface theories, is now seen to couple curvature gradients and moments.\textspace

Given an appropriate constitutive model and the electric field, Eqs.~\eqref{eq:inplane_eq_1} and~\eqref{eq:shape_eq_1} only depend on the density $\rho_{\mathrm{s}}$, velocity $\bm{v}_0$, and the stress component $S^\Ga$. According to Eq.~\eqref{eq:Talpha_n_general}, $S^\Ga = \delta\bmT_0^\Ga\cdot \bmn$ is determined by the reactive stresses and therefore requires an additional equation to solve for it. That equation is obtained from the first-order linear and zeroth-order angular momentum balances. To this end, first note that the in-plane components of the first-order stress vector divergence can be expressed as
\begin{align}
    -\frac{\delta^2}{4} \bm{T}^\alpha_{1:\alpha} \cdot \bm{a}^\beta =  M^{\gamma \beta}_{;\gamma} - R^\gamma b_\gamma^\beta~, \label{eq:Maa_inplane}
\end{align}
which, upon substitution of the first-order linear momentum balance in Eq.~\eqref{eq:1storder_linmom}, reduces the zeroth-order angular momentum in Eq.~\eqref{eq:inplane_0_angmom} to
\begin{align}
    S^\alpha &= -M^{\gamma\alpha}_{;\gamma} + \delta \left(\frac{1}{2}\left(\bm{t}^+ - \bm{t}^-\right) + \left\langle \left\llbracket \bm{\sigma}^T_\mathrm{M} \right\rrbracket \right\rangle^\mathcal{M} \bm{n}\right) \cdot \bm{a}^\alpha \nonumber \\
    &\hspace{2cm} - \frac{\delta}{4} \left(  \rho_{\mathrm{s}}\delta H \dot{\bm{v}}_0 +  \frac{\rho_{\mathrm{s}}\delta}{2} \ddot{\bm{n}}  + 2\delta H\left(\frac{1}{2}\left(\bm{t}^+ + \bm{t}^-\right) + \frac{1}{2}\left\llbracket \left\llbracket \bm{\sigma}^T_\mathrm{M} \right\rrbracket \right\rrbracket^\mathcal{M} \bm{n}\right)  - \delta\bm{f}_1 \right)\cdot \bma^\Ga~. \label{eq:Sa_final}
\end{align}
In Eq.~\eqref{eq:Sa_final}, we neglected the term $\frac{\delta^3}{4} \left(2H^2-K\right)\bmT_0^3 \cdot \bma^\Ga$ as it is small compared to others in Eq.~\eqref{eq:Sa_final} based on Eqs.~\eqref{eq:H2small} and~\eqref{eq:Ksmall}. Additionally, we omitted the term $\frac{\delta^3}{4} H_{,\Gb} \left( \bmT_0^\Gb + \frac{1}{2} \bmT_2^\Gb\right) \cdot \bma^\Ga$ as it is small upon substitution of Eq.~\eqref{eq:Sa_final} into Eqs.~\eqref{eq:inplane_eq_1}  and~\eqref{eq:shape_eq_1} based on Eqs.~\eqref{eq:ka_ll1},~\eqref{eq:lcls}--\eqref{eq:ells_ll1},~\eqref{eq:stress_assumption_1}, and~\eqref{eq:stress_assumption_2}. Furthermore, the term\footnote{Note that we are not including the body force term $\delta \bmf_1$ in Eq.~\eqref{eq:Sa_inplane_neglect}.}
\begin{align}
    - \frac{\delta}{4} \left(  \rho_{\mathrm{s}}\delta H \dot{\bm{v}}_0 +  \frac{\rho_{\mathrm{s}}\delta}{2} \ddot{\bm{n}}  + 2\delta H\left(\frac{1}{2}\left(\bm{t}^+ + \bm{t}^-\right) + \frac{1}{2}\left\llbracket \left\llbracket \bm{\sigma}^T_\mathrm{M} \right\rrbracket \right\rrbracket^\mathcal{M} \bm{n}\right) \right)\cdot \bma^\Ga~, \label{eq:Sa_inplane_neglect}
\end{align}
appearing in Eq.~\eqref{eq:Sa_final}, is negligible when substituting it into the in-plane components of the linear momentum balance in Eq.~\eqref{eq:inplane_eq_1}, as can be seen by invoking Eqs.~\eqref{eq:H2small},~\eqref{eq:cbaa},~\eqref{eq:va_1_magnitude} and~\eqref{eq:ddotn}. Note, however, that the term in Eq.~\eqref{eq:Sa_inplane_neglect} cannot be neglected when considering the normal component of the linear momentum balance in Eq.~\eqref{eq:shape_eq_1}.\textspace 

Equation~\eqref{eq:Sa_final} provides the equation required to solve for $S^\Ga$ in the $(2+\delta)$-dimensional theory. In strict surface theories, $S^\Ga$ takes the role of a transverse shear stress that must be balanced by moment changes to satisfy the angular momentum balance, i.e. $S^\alpha = -M^{\gamma\alpha}_{;\gamma}$. In Eq.~\eqref{eq:Sa_final}, we find several additional contributions arising from inertial terms and external tractions.  While the inertial terms are typically negligible for lipid membranes \cite{sahu2020geometry}, the tractions due to the coupling between lipid membranes and their surrounding bulk fluid may play a significant role \cite{sens2002undulation,lacoste2009electrostatic,abreu2014fluid,shibly2016experimental,vlahovska2019electrohydrodynamics}.\textspace

In summary, the $(2+\delta)$-dimensional equations of motion are given by the mass balance in Eq.~\eqref{eq:mass_zeroth_truncated_final} and Eqs.~\eqref{eq:inplane_eq_1},~\eqref{eq:shape_eq_1}, and~\eqref{eq:Sa_final}. When used in conjunction with a constitutive model, these equations determine the fundamental unknowns $\rho_\mathrm{s}$ and $\bmv_0$.

\section{Conclusion}

Continuum theories treating lipid membranes as strictly two-dimensional surfaces have limitations in capturing their coupled electrical and mechanical behavior. These theories fail to resolve surface charges at the bulk-membrane interfaces, the potential drop across the membrane, and the electric field within the membrane. Therefore, strict surface theories also fail to capture the effects of Maxwell stresses within the membrane and its vicinity. Additionally, strict surface theories cannot incorporate the distinct traction and velocity continuity conditions at the two interfaces between lipid membranes and their surrounding fluids.\textspace

To avoid the aforementioned shortcomings, we employ the dimension reduction method proposed in part 1 \cite{omar2023ES} to derive an effective surface theory for the electromechanics of lipid membranes without considering the limit of vanishing thickness. Starting from the three-dimensional mass, linear, and angular momentum balances and the Kirchhoff-Love assumptions, we obtain the $(2+\delta)$-dimensional balance laws, which do not suffer from the shortcomings of strict surface theories. These yield the $(2+\delta)$-dimensional equations of motion for the unknown mid-surface velocity and density. These balance laws and equations of motion are general and not specific to any choice of constitutive model.\textspace

The generality of the theory presented in this article suggests it does not only hold relevance for the electromechanics of lipid membranes but also for the broader field of thin shell theory \cite{kirchhoff1850gleichgewicht,ventsel2001thin,ugural2009stresses,altenbach2017thin}. The historical and contemporary developments in thin shell theory can be broadly categorized into two distinct approaches. The first is to treat the body as a two-dimensional surface embedded in a three-dimensional space and propose new balance laws and constitutive models for such a surface, resulting in strict surface theories (e.g. \cite{ericksen1957exact,sanders1963nonlinear,naghdi1973theory,steigmann1999relationship,simo1989stress}). The second approach is to consider a three-dimensional body and deduce effective balance laws and constitutive models from their three-dimensional counterparts (e.g. \cite{love1927treatise,chien1944intrinsic,zerna1949beitrag,naghdi1962foundations,koiter1967stability,naghdi1973theory,ciarlet1979justification,libai1998nonlinear}). The method presented in this article belongs to the latter category, which will be the focus of the following discussion. There exists a number of methods used to derive the governing equations of thin shells from a three-dimensional starting point, such as asymptotic expansions (e.g. \cite{chien1944intrinsic,koiter1967stability,miara1994justificationI,miara1994justificationII,steigmann2013koiter}) and $\Gamma$-convergence \cite{percivale2008introduction, braides2002gamma}. While an extensive discussion of existing shell theories is beyond the scope of this article, we note that only a few approaches hold for arbitrary material models and deformations. In the following, we compare the theory developed in this article only to the popular theories by Podio-Guidugli \cite{podio1989exact,podio1990constraint} and Naghdi \cite{naghdi1973theory} as they share similarities with the method proposed in this article.\textspace

Podio-Guidugli introduced the so-called method of internal constraints to reduce the three-dimensional balance laws of thin elastic bodies to effective surface equations. In Ref.~\cite{podio1989exact}, Podio-Guidugli realizes the inconsistency introduced by restricting the kinematics to satisfy the \KL assumptions while using a constitutive model that is agnostic to any such restriction. This inconsistency is circumvented by viewing the \KL assumptions as constraints and introducing corresponding reactive stresses similar to the approach taken in this article. Podio-Guidugli linearizes the strain measure and chooses a linear constitutive model, which permits solving for the reactive stresses and explicitly integrating the equations of motion through the thickness to obtain an effective surface theory. For nonlinear problems, however, the reactive stresses cannot be solved for a-priori, and integration through the thickness is challenging. Thus, while the method of internal constraints has been developed further in Refs.~\cite{podio1990constraint,podio2000recent,podio2000internal,favata2012new,podio2017six}, it currently remains restricted to linear problems. The $(2+\delta)$-dimensional theory developed here also treats the \KL assumptions as constraints but is not restricted to small deformations and linear constitutive models. 
This comes at the cost of having to assume that reactive stresses of second and higher polynomial order are small, allowing us to determine the reactive stresses from the traction continuity conditions.
Lastly, we note that integration through the thickness, as done in the method of internal constraints, implies that the balance laws are satisfied in a global rather than a pointwise sense along the thickness. In contrast, the spectral approach used to derive the $(2+\delta)$-dimensional theory implies a pointwise solution of the balance laws.\textspace

The thin shell theory originally introduced by Naghdi and Nordgren \cite{naghdi1963nonlinear} and later generalized by Naghdi \cite{naghdi1973theory} proposes the definition of so-called stress resultants as new unknowns, together with the derivation of additional balance laws. This permits solving for the $M$ coefficients $\bmx_n$ of a position vector of the form $\bmx = \sum_{n=0}^{M-1} \bmx_n \left(\xi^3\right)^n$. The new balance laws required for this approach are found by multiplying the energy balance by $\left(\xi^3\right)^n$ and invoking the Green-Naghdi-Rivlin (GNR) theorem\footnote{A similar approach is presented in Ref.~\cite{podio2000recent} where the principle of virtual work is invoked and the multiplication by $\left(\xi^3\right)^n$ takes on the role of test functions to generate additional equations.} \cite{green1964cauchy}. This motivates the definition of the stress resultants $\hat{\bmN}^{(n) \Ga}$ in terms of the three-dimensional stress vectors as 
\begin{align}
    \hat{J}_0 \hat{\bmN}^{(n) \Ga} = \int_{-\delta/2}^{\delta/2} \hat{\bmT}^\Ga \left(\xi^3\right)^n \diff{\hat{s}}~, \quad n = 0,...,M~.\label{eq:stress_resultant}
\end{align}
The stress resultants could be determined from the three-dimensional constitutive models. In practice, however, evaluation of the integral in Eq.~\eqref{eq:stress_resultant} is challenging and usually restricted to the linear case. Instead, one must propose new constitutive models relating the stress vectors $\hat{\bmN}^{(n) \Ga}$ to the position vector coefficients $\bmx_k$ using, for instance, the Coleman-Noll procedure \cite{naghdi1973theory} or linear irreversible thermodynamics \cite{sahu2017irreversible}. When choosing $M=1$ in the expansion of the position vector above, the approach proposed in this article shares similarities with Naghdi’s approach. First, we note that we could also start the derivations of the $(2+\delta)$-dimensional theory from the energy balance, take inner products with the Chebyshev polynomials, and subsequently invoke the GNR theorem. However, since only rigid body motions are considered in the GNR theorem, this approach is equivalent to starting from the linear and angular momentum balances directly. Second, instead of multiplying the balance laws by $\left(\xi^3\right)^n$ before integrating, we multiply by the Chebyshev polynomials $\ChebTR{k}$ and the weight $1/\sqrt{1-\Theta^2}$ (see Eqs.~\eqref{eq:InnerProd_def} and~\eqref{eq:inner_prod_eqs}). The use of orthogonal polynomials simplifies the evaluation of the integrals, eliminating the need to consider the expansion coefficients as new unknowns and finding constitutive models for them. Hence, three-dimensional constitutive models can be directly substituted into the $(2+\delta)$-dimensional equations of motion, leading to a dimensionally-reduced theory that is consistent with the three-dimensional starting point. This procedure is presented in detail in part 3 of this series of articles, where we propose constitutive models to specialize the $(2+\delta)$-dimensional theory to the electromechanics of lipid membranes.

\section*{Acknowledgements}
Y.A.D.O. and K.K.M. were entirely supported by the Director, Office of Science, Office of Basic Energy Sciences, of the US Department of Energy under Contract No. DEAC02-05CH1123. The authors would like to thank Dr. Dimitrios Fraggedakis for insightful discussions, in particular on the question of reactive stresses, and Ahmad Alkadri for guidance on the derivation of the contravariant basis vectors. They also express their appreciation to Dr. Amaresh Sahu and Yulong (Lewis) Pan for valuable insights regarding the material time derivative of tangent vectors.

%% file: electromechanics-article/Figures/thin_shell_setup.tex
\tikzset{>=latex}
\begin{tikzpicture}[font=\small, x=0.8\linewidth, y=0.8\linewidth] 

\pgfdeclarelayer{background}
\pgfdeclarelayer{foreground}
\pgfsetlayers{background,foreground}

\def \h{0.07}

\begin{pgfonlayer}{background}
\coordinate (A) at (0,0.1); 
\coordinate (B) at (0.6,0.1); 
\coordinate (C) at (1,0.3); 
\coordinate (D) at (0.4,0.3); 
\coordinate (topcenter) at (0.55, 0.28);
\coordinate (bottomcenter) at (0.55, 0.109);
\coordinate (nustart) at (0.96, 0.3);

\draw[semithick] 
      (A) .. controls +(0.2, -0.05) and +(-0.25,0.05)  .. (B)
      (B) .. controls +(0.15, 0.15) and +(-0.15, -0.15) .. (C)
      (A) -- ++(0,\h) coordinate (E)
      (B) -- ++(0,\h) coordinate (F)
      (C) -- ++(0,\h) coordinate (G)
      (D) -- ++(0,\h) coordinate (H);
      
\end{pgfonlayer}

\begin{pgfonlayer}{foreground}
\filldraw[fill=white, semithick] 
      (E) .. controls +(0.2, -0.05) and +(-0.25,0.05)  .. (F) .. controls +(0.15, 0.15) and +(-0.15, -0.15) .. (G) .. controls +(-0.2, -0.05) and +(0.2, 0.05) .. (H) .. controls +(-0.15, -0.1) and +(0.15, 0.1) .. (E);

\node [position=-7pt:7pt from A] (hnode) {$\delta$};
\node [position=-45pt:-5pt from topcenter] (ltop) {};
\node [position=20pt:15pt from B] (rbot) {};
\node [position=32pt:15.5pt from rbot] (rmid) {};
\node [position=42pt:2pt from A] (fmid) {};

\draw[] (ltop) .. controls +(-0.1,0) and +(0.1,-0.) .. ++(-0.2,0.08) node[anchor=east] (nodeSp) {$\mathcal{S}^+$};
\draw (rbot) .. controls +(0.07,0) and +(-0.07,-0.) .. ++(0.15,-0.06) node[anchor=west] (nodeSm) {$\mathcal{S}^-$}; 
\draw (rmid) .. controls +(0.07,0) and +(-0.07,-0.) .. ++(0.15,-0.06) node[anchor=west] (nodeS0) {$\mathcal{S}_0$}; 
\draw (fmid) .. controls +(-0.04,0) and +(0.04,0) .. +(-0.1, -0.1) node[anchor=east] (nodeSpar) {$\mathcal{S}_{||}$}; 
\node[position=80pt:32pt from A] (nodeM) {$\mathcal{M}$};
\node[position=-70pt:35pt from C] (nodeBp) {$\mathcal{B}^+$};
\node[position=-50pt:-20pt from B] (nodeBm) {$\mathcal{B}^-$};

\draw[-{Stealth}, thick] (topcenter) -- ++(-0.01, 0.15);
\node [position=10pt:10pt from topcenter] (nplabel) {$\bar{\bm{n}}^+$};  
\draw[-{Stealth}, thick] (bottomcenter) -- ++(0.003, -0.08);
\node [position=15pt:-10pt from bottomcenter] (nmlabel) {$\bar{\bm{n}}^-$};  

\draw[-{Stealth}, thick] (nustart) -- ++(0.15, 0.);
\node [position=30pt:10pt from nustart] (nulabel) {$\bm{\nu}$};  

\draw[dashed] 
    ($(A)+(0.,\h/2)$) .. controls +(0.2, -0.05) and +(-0.25,0.05)  .. ($(B)+(0,\h/2)$)
    ($(B)+(0.,\h/2)$) .. controls +(0.15, 0.15) and +(-0.15, -0.15) .. ($(C)+(0.,\h/2)$);
\end{pgfonlayer}

\end{tikzpicture}

%% file: electromechanics-article/Figures/S0_parametrization.tex
\begin{tikzpicture}[font=\small, x=0.8\linewidth, y=0.8\linewidth] 
\node(CScenter) at (0.33,0)  {};
\draw[semithick,->] (CScenter.center) -- +(0,0.6) coordinate (ez);
\draw[semithick,->] (CScenter.center) -- +(2/3,0) coordinate (ey);
\draw[semithick,->] (CScenter.center) -- +(-.5/3,-.5/3) coordinate (ex);

\coordinate (A) at (0.1,0.2); 
\coordinate (B) at (0.7,0.2); 
\coordinate (C) at (1.1,0.4); 
\coordinate (D) at (0.5,0.4); 

\begin{scope}
\draw[smooth, fill=gray,fill opacity=.3] (0.2,0.15) coordinate (BL) .. controls +(0.05,0.1) and +(-0.05,-0.06) ..  +(0.3,0.3) .. controls +(0.3,0.05) and +(-0.3,-0.03) .. +(1,0.3)  .. controls +(-0.15,-0.1) and +(0.2,0.15) .. +(0.7,0)  .. controls +(-0.3,0.1) and +(0.33,-0.033) .. (BL) ;
\end{scope}

\draw[dotted, semithick] (0.28,0.25) .. controls +(0.1,-0.05) and +(-0.1,-0.05) .. (0.7,0.3) coordinate (midcross) .. controls +(0.1,0.05) and +(-0.1,0) .. (1.09,0.35);
\draw[dotted, semithick] (0.6,0.18) .. controls +(0.1,0.08) and +(-0.01,-0.03) .. (0.7,0.3) .. controls +(0.01,0.03) and +(0.05,-0.05) .. (0.65,0.47);  


 \draw[->,>=stealth, dashed, thick,shorten <=3] (0.5396,0.17) -- (0.7,0.3);
 \draw[thick] (CScenter.center) -- (0.5396,0.17);

\node[] at (1,0.4) {$\mathcal{S}_0$};
\node[] at (0.41,0.28) {$\xi^1$};
\node[] at (0.63,0.41) {$\xi^2$};
\node[position=15pt:0pt from ex.east] (nodeex) {$\bm{e}_x$};
\node[position=0pt:-10pt from ey] (nodeey) {$\bm{e}_y$};
\node[position=-10pt:0pt from ez] (nodeez) {$\bm{e}_z$};
\node[] at (0.48,0.06) {$\hat{\bm{x}}_0$};

\end{tikzpicture}

%% file: electromechanics-article/Figures/meshing_thickness.tikz
\begin{tikzpicture}[font=\normalfont, x=\textwidth, y=0.65\textwidth]

\useasboundingbox (0,0) rectangle (1,1);

\newcommand\CPix{0.4}
\newcommand\CPiy{-0.15}
\newcommand\CPiix{-0.4}
\newcommand\CPiiy{0.15}
\newcommand\SNcenter{0.5}

\coordinate (S0left) at (0,0.45);
\coordinate (S0right) at (1,0.55);
\coordinate (Spleft) at (0,0.54);
\coordinate (Spright) at (1,0.64);
\coordinate (Smleft) at (0,0.36);
\coordinate (Smright) at (1,0.46);

\coordinate (Sm1left) at (0,0.24);
\coordinate (Sm1right) at (1,0.34);
\coordinate (Sm2left) at (0,0.07);
\coordinate (Sm2right) at (1,0.17);
\coordinate (Sp1left) at (0,0.68);
\coordinate (Sp1right) at (1,0.78);
\coordinate (Sp2left) at (0,0.83);
\coordinate (Sp2right) at (1,0.93);

\newcommand\OnP{4pt}
\newcommand\OffP{3pt}

\draw[smooth, thick] (Spleft) .. controls +(\CPix, \CPiy) and +(\CPiix, \CPiiy) .. (Spright);
\draw[smooth, thick] (Smleft) .. controls +(\CPix, \CPiy) and +(\CPiix, \CPiiy) .. (Smright);
\draw[smooth, very thick, dashed, dash pattern = on 2pt off 2pt] (S0left) .. controls +(\CPix, \CPiy) and +(\CPiix, \CPiiy) .. (S0right);

\begin{scope}
\draw[smooth, fill=gray,fill opacity=.3,draw=none]  (Spleft) .. controls +(\CPix, \CPiy) and +(\CPiix, \CPiiy) .. (Spright) .. controls (S0right) .. (Smright) .. controls +(\CPiix, \CPiiy) and +(\CPix,\CPiy) .. (Smleft) ;
\end{scope}
\begin{pgfonlayer}{bg1}
\begin{scope}
    \draw[smooth, fill=white,draw=none]  (Spleft) .. controls +(\CPix, \CPiy) and +(\CPiix, \CPiiy) .. (Spright) .. controls (S0right) .. (Smright) .. controls +(\CPiix, \CPiiy) and +(\CPix,\CPiy) .. (Smleft) ;
\end{scope}
\end{pgfonlayer}

\draw[smooth, thick, dashed, dash pattern= on \OnP off \OffP] (Sm1left) .. controls +(\CPix, \CPiy) and +(\CPiix, \CPiiy) .. (Sm1right);
\draw[smooth, thick, dashed, dash pattern= on \OnP off \OffP] (Sm2left) .. controls +(\CPix, \CPiy) and +(\CPiix, \CPiiy) .. (Sm2right);
\begin{pgfonlayer}{bg0}
    \draw[thick, dashed,dash phase=0pt, dash pattern= on \OnP off \OffP] (0.1, 0.5) -- +(-0.05,-0.5);
    \draw[thick, dashed,dash phase=0pt, dash pattern= on \OnP off \OffP] (0.25, 0.5) -- +(0.07,-0.5);
    \draw[thick, dashed, dash phase=3pt, dash pattern= on \OnP off \OffP] (0.48, 0.5) -- +(0.07,-0.5);
    \draw[thick, dashed,dash phase=4pt, dash pattern= on \OnP off \OffP] (0.7, 0.5) -- +(0.04,-0.5);
    \draw[thick, dashed,dash phase=4pt, dash pattern= on \OnP off \OffP] (0.95, 0.5) -- +(-0.05,-0.5);
\end{pgfonlayer}

\draw[smooth, thick, dashed, dash pattern= on \OnP off \OffP] (Sp1left) .. controls +(\CPix, \CPiy) and +(\CPiix, \CPiiy) .. (Sp1right);
\draw[smooth, thick, dashed, dash pattern= on \OnP off \OffP] (Sp2left) .. controls +(\CPix, \CPiy) and +(\CPiix, \CPiiy) .. (Sp2right);
\begin{pgfonlayer}{bg0}
    \draw[thick, dashed,dash phase=4pt, dash pattern= on \OnP off \OffP] (0.05, 0.5) -- +(0.05,0.5);
    \draw[thick, dashed,dash phase=4pt, dash pattern= on \OnP off \OffP] (0.3, 0.5) -- +(-0.03,0.5);
    \draw[thick, dashed,dash phase=4pt, dash pattern= on \OnP off \OffP] (0.49, 0.5) -- +(-0.04,0.5);
    \draw[thick, dashed,dash phase=1pt, dash pattern= on \OnP off \OffP] (0.7, 0.5) -- +(-0.05,0.5);
    \draw[thick, dashed,dash phase=0pt, dash pattern= on \OnP off \OffP] (0.85, 0.5) -- +(0.04,0.5);
\end{pgfonlayer}

\node[] at (0.18,0.58) {$\mathcal{S}^+$};
\node[] at (0.2,0.27) {$\mathcal{S}^-$};
\draw (0.31,0.43) .. controls +(0.0,0.05) and +(-0.05,0) .. ++(0.06,0.19) node[anchor=west, outer sep  = 0, inner sep =0.2em] (nodeSp) {$\mathcal{S}_0$}; 

\end{tikzpicture}

%% file: electromechanics-article/Figures/meshing_nothickness.tikz
\begin{tikzpicture}[font=\normalfont, x=\textwidth, y=0.8\textwidth]

\useasboundingbox (0,0) rectangle (1,1);

\coordinate (S0left) at (0,0.45);
\coordinate (S0right) at (1,0.55);

\coordinate (Sm1left) at (0,0.33);
\coordinate (Sm1right) at (1,0.43);
\coordinate (Sm2left) at (0,0.16);
\coordinate (Sm2right) at (1,0.26);
\coordinate (Sp1left) at (0,0.59);
\coordinate (Sp1right) at (1,0.69);
\coordinate (Sp2left) at (0,0.74);
\coordinate (Sp2right) at (1,0.84);

\newcommand\CPix{0.4}
\newcommand\CPiy{-0.15}
\newcommand\CPiix{-0.4}
\newcommand\CPiiy{0.15}
\newcommand\SNcenter{0.5}

\coordinate (mesh1) at (0.1, 0.42);
\coordinate (mesh2) at (0.25, 0.42);
\coordinate (mesh3) at (0.48, 0.49);
\coordinate (mesh4) at (0.7, 0.57);
\coordinate (mesh5) at (0.95, 0.565);
\newcommand\lowb{0.09}
\newcommand\upb{0.91}

\newcommand\OnP{4pt}
\newcommand\OffP{3pt}

\draw[smooth, very thick, dashed, dash pattern = on 2pt off 2pt] (S0left) .. controls +(\CPix, \CPiy) and +(\CPiix, \CPiiy) .. (S0right);

\draw[smooth, thick, dashed, dash pattern= on \OnP off \OffP] (Sm1left) .. controls +(\CPix, \CPiy) and +(\CPiix, \CPiiy) .. (Sm1right);
\draw[smooth, thick, dashed, dash pattern= on \OnP off \OffP] (Sm2left) .. controls +(\CPix, \CPiy) and +(\CPiix, \CPiiy) .. (Sm2right);

\draw[thick, dashed, dash pattern= on \OnP off \OffP] (mesh1) -- (0.05,\lowb);
\draw[thick, dashed, dash pattern= on \OnP off \OffP] (mesh2) -- (0.32,\lowb);
\draw[thick, dashed, dash pattern= on \OnP off \OffP] (mesh3) -- (0.55,\lowb);
\draw[thick, dashed, dash pattern= on \OnP off \OffP] (mesh4) -- (0.74,\lowb);
\draw[thick, dashed, dash pattern= on \OnP off \OffP] (mesh5) -- (1,\lowb);

\draw[smooth, thick, dashed, dash pattern= on \OnP off \OffP] (Sp1left) .. controls +(\CPix, \CPiy) and +(\CPiix, \CPiiy) .. (Sp1right);
\draw[smooth, thick, dashed, dash pattern= on \OnP off \OffP] (Sp2left) .. controls +(\CPix, \CPiy) and +(\CPiix, \CPiiy) .. (Sp2right);
\begin{pgfonlayer}{bg1}
    \draw[thick, dashed, dash pattern= on \OnP off \OffP] (mesh1) -- (0.05,\upb);
    \draw[thick, dashed, dash pattern= on \OnP off \OffP] (mesh2) -- (0.26,\upb);
    \draw[thick, dashed, dash pattern= on \OnP off \OffP] (mesh3) -- (0.44,\upb);
    \draw[thick, dashed, dash pattern= on \OnP off \OffP] (mesh4) -- (0.65,\upb);
    \draw[thick, dashed, dash pattern= on \OnP off \OffP] (mesh5) -- (0.99,\upb);
\end{pgfonlayer}

\node[] at (0.32,0.49) {$\mathcal{S}_0$};

\end{tikzpicture}